\documentclass[12pt]{iopart}

\usepackage{graphicx}
\usepackage{bm}

\usepackage{mathptmx, courier, pifont}
\usepackage[scaled=0.92]{helvet}
\usepackage[T1]{fontenc}
\usepackage{textcomp}

\newlength{\figdn}
\newcommand{\Fplus}[1]{F^+_{#1}}
\newcommand{\Fminus}[1]{F^-_{#1}}
\newcommand{\fplus}[1]{f^+_{#1}}
\newcommand{\fminus}[1]{f^-_{#1}}
\newcommand{\PEsource}[2]{f_{{#1}{\rightleftharpoons}{#2}}}
\newcommand{\eq}[1]{Eq.~(\ref{#1})}
\newcommand{\eqnoeq}[1]{(\ref{#1})}
\newcommand{\fig}[1]{Fig.~\ref{fig:#1}}
\newcommand{\diffelement}[1]{d#1}
\newcommand{\deriv}[2]{\frac{d#1}{d#2}}
\newcommand{\pardiv}[2]{\frac{\partial{#1}}{\partial{#2}}}
\newcommand{\tsub}[1]{_{\mbox{\scriptsize#1}}}

\newcommand{\tfrac}[2]{\mbox{$\small\frac{#1}{#2}$}}
\newcommand{\units}[1]{\mbox{\ #1}}
\newcommand{\isotope}[2]{\mbox{$^{#1}$#2}}

\newcommand{\mwgalign}[1]{\hskip-0.0em&#1&\hskip-0.0em}
\newcommand{\flush}{&&}
\newcommand{\kforward}[1]{k^{(#1)}_{\rm\scriptstyle f}}
\newcommand{\kback}[1]{k^{(#1)}_{\rm\scriptstyle r}}

\newcommand{\putfig}[6]{%
\begin{figure}\vspace*{#2}%
\begin{flushright}%
\includegraphics*[scale=#4]{#5}%
\end{flushright}%
\caption{#6}%
\vspace*{#3}%
\label{fig:#1}%
\end{figure}}

\begin{document}

\title
{Explicit Integration of Extremely-Stiff Reaction
Networks: Partial Equilibrium Methods}

\author{M. W. Guidry$^{1,2,3}$, J. J. Billings$^3$, and W. R. Hix$^{1,2,3}$}

\address{$^1$Department of Physics and Astronomy, University of Tennessee,
Knoxville, TN 37996-1200, USA}
\address{$^2$Physics Division, Oak Ridge National Laboratory, Oak Ridge, TN
37830, USA}
\address{$^3$Computer Science and Mathematics Division, Oak Ridge National
Laboratory, Oak Ridge, TN 37830, USA}

\ead{guidry@utk.edu}

\begin{abstract}
In two preceding papers \cite{guidAsy,guidQSS} we have shown that, when reaction
networks are well-removed from equilibrium, explicit asymptotic and
quasi-steady-state approximations can  give algebraically-stabilized integration
schemes that rival standard implicit methods in accuracy and speed for extremely
stiff systems. However, we also showed that these explicit methods remain
accurate but are no longer competitive in speed as the network approaches
equilibrium. In this paper we analyze this failure and show that it is
associated with the presence of fast equilibration timescales that neither
asymptotic nor quasi-steady-state approximations are able to remove efficiently
from the numerical integration. Based on this understanding, we develop a
partial equilibrium method to deal effectively with the approach to equilibrium
and show that explicit asymptotic methods, combined with the new partial
equilibrium methods, give an integration scheme that plausibly can deal with the
stiffest networks, even in the approach to equilibrium, with accuracy and speed
competitive with that of implicit methods. Thus we demonstrate that such
explicit methods may offer alternatives to implicit integration of even
extremely stiff systems, and that these methods may permit integration of much
larger networks than have been possible before in a number of fields. 
\end{abstract}


\vspace{2pc}



\section{\label{intro} Introduction}

Problems from many fields of science and technology require the solution of
large coupled reaction networks describing the flow of population between
various sources and sinks.  Some important examples include  reaction networks
in combustion chemistry \cite{oran05}, geochemical cycling of elements
\cite{magick}, and thermonuclear reaction networks in astrophysics
\cite{hix05,timmes}.  The systems of differential equations that are commonly
used to model these reaction networks usually exhibit stiffness, which we may
think of loosely as arising from multiple timescales in the problem that differ
by many orders of magnitude \cite{oran05,gear71,lamb91,press92}.  The most
straightforward way to solve such a system might appear to be an explicit
numerical integration scheme (for an explicit method, advancing a timestep
requires only information known from previous timesteps).  However, the
textbooks routinely state \cite{oran05,lamb91,press92} that such systems cannot
be integrated efficiently using explicit forward finite-difference methods
because of stability issues:  for an explicit algorithm, the maximum stable
timestep in a stiff system is typically set by the fastest timescales, even if
those timescales are not of central interest. The traditional solution of the
stiffness problem is to replace explicit integration with implicit integration
(integration methods that require the values for derivatives at timesteps not
yet evaluated).

The astrophysical CNO cycle for conversion of hydrogen to helium
(Fig.~\ref{fig:cnoCycle}) is an instructive example of the stiffness issue. In
the CNO cycle the fastest rates typically are $\beta$-decays with half-lives of
order 100 seconds, but tracking main-sequence hydrogen burning may require
integration of the hydrogen-burning network for as long as billions of years
($\sim 10^{16}$ seconds). With explicit forward differencing the largest stable
integration timestep is set by the fastest rates and will be of order $10^2$
seconds, so $\sim 10^{14}$ explicit integration steps could be required, even
for the idealized case of constant temperature and density.  Conversely, this
same numerical integration requires at most a few hundred steps using implicit
methods. For this reason, it is generally thought that explicit methods are not
viable for extremely stiff networks. To quote {\em Numerical Recipes}
\cite{press92}, ``For stiff problems we {\em must} use an implicit method if we
want to avoid having tiny stepsizes.'' 
\putfig
{cnoCycle}
{0pt}
{\figdn}
{0.90}
{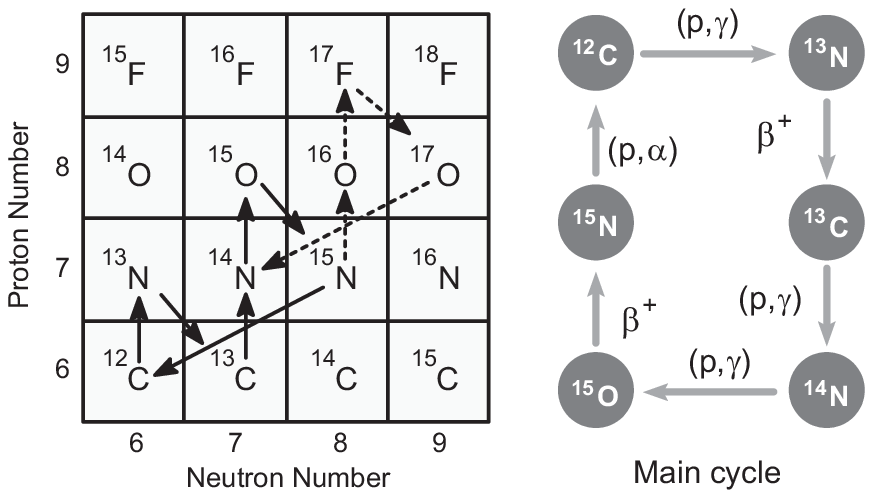}
{The CNO cycle. On the left side the main branch of the cycle is illustrated
with solid arrows and a side branch is illustrated with dashed arrows. On the
right side, the main branch of the CNO cycle is illustrated in more detail
($p$ stands for \isotope1H and $\alpha$ for \isotope 4{He}).}

Although implicit methods typically are stable for stiff systems, they require
substantial additional computational overhead relative to explicit methods.  For
large coupled sets of equations, this normally takes the form of iterative
solutions that require the inversion of large matrices at each step. The
required matrix inversions dictate that, except where simplifications based on
matrix structure can be exploited (for example, sparse-matrix methods), implicit
algorithms may scale as poorly as quadratically or cubically with the size of
the network. Thus, implicit methods can be expensive to implement for large
networks, particularly those that are coupled in real time to a broader problem
such as hydrodynamical evolution.

Despite the generally negative view of explicit methods for stiff systems
sketched in the preceding paragraphs, they would be attractive options for
complex networks if they could take larger timesteps because of their overall
simplicity and highly-favorable linear scaling with network size relative to
implicit methods. To increase the integration step size for explicit methods
obviously requires overcoming formidable numerical stability problems that are
associated with the stiffness. In principle, this might be accomplished by using
approximate analytical solutions to reduce the stiffness of the equation set to
be solved, thereby improving the overall stability of the network. In the first
two papers of this series \cite{guidAsy,guidQSS}, we have shown in a variety of
examples that this can be accomplished for systems that are not near equilibrium
by using asymptotic and quasi-steady-state approximations to stabilize the
numerical integration. However, as we shall now discuss, the nature of stiffness
is different for networks that are near equilibrium. This implies that the
modifications required to stabilize standard explicit methods when stiffness is
encountered far from equilibrium must be altered when that same network
approaches equilibrium.

\section{\label{sh:varieties} Varieties of Stiffness}

For the large and very stiff networks that are our primary interest here there
are several fundamentally different sources of stiffness instability.  Most
textbook discussions of stiffness concentrate on an instability associated with
small quantities that strictly should be non-negative being driven negative
by an overly-ambitious numerical integration step. However, there are other
guises that stiffness can take, as we now discuss. The equations that we must
integrate take the general form
 \begin{eqnarray}
\deriv{y_i}{t} &=&
\Fplus{i} - \Fminus{i} 
\nonumber
\\
&=& (\fplus 1 + \fplus 2 + \ldots)_i  - (\fminus 1 + \fminus 2 + \ldots)_i
\nonumber
\\
&=& (\fplus 1 - \fminus 1)_i + (\fplus 2 - \fminus 2)_i + \ldots 
 = \sum_j (\fplus j - \fminus j)_i,
\label{equilDecomposition}
\end{eqnarray}
where the $y_i (i=1 \dots N)$ describe the dependent variables (abundance in our
examples), $t$ is the independent variable (time in our examples), the fluxes
between species $i$ and $j$ are denoted by $(f^{\pm}_j)_i$, and the sum for each
variable $i$ is over all variables $j$ coupled to $i$ by a non-zero flux
$(f^{\pm}_{j})_i$. For an $N$-species network there will be $N$ such equations
in the populations $y_i$, generally coupled to each other because of the
dependence of the fluxes on the different $y_j$.

Because our discussion is intended to be quite general, we shall most often
formulate the equations \eqnoeq{equilDecomposition} in terms of generic
population variables $y_i$ that are assumed to be proportional to the number
density of species $i$. Where we give specific results for astrophysical
networks, we shall use population variables most common for that field, such as 
the mass fraction $X_i$ and the (molar) abundance $Y_i$, with
\begin{equation}
X_i  = \frac{n_iA_i}{\rho N\tsub A}
\qquad
Y_i \equiv \frac{X_i}{A_i} = \frac{n_i}{\rho N\tsub A}
\label{5.35}
\end{equation}
where $N\tsub A$ is Avogadro's number, $\rho$ is the total mass density, $A_i$
is the atomic mass number and $n_i$ the number density for the species $i$, and
by definition the mass fractions sum to unity if nucleon number is conserved:
$\sum X_i =1$.

In \eq{equilDecomposition} the total flux has been decomposed into  a component
$\Fplus i$ that increases the population $y_i$ and a component $\Fminus i$ that
depletes the population $y_i$, and in the third line this has been decomposed
further into individual groups of terms  $\fplus j-\fminus j$. In the approach
to equilibrium a species population entails a delicate balance between a total
flux $\Fplus i$ populating the species and a total flux $\Fminus i$ depleting
it. Near equilibrium the difference $F_i=\Fplus i-\Fminus i$ can be orders of
magnitude smaller than $\Fplus i$ or $\Fminus i$ and small numerical errors in
$\Fplus i$ or $\Fminus i$ can produce large errors in the difference $F_i$.
Because of the population coupling in complex networks, this error propagates
and compromises the accuracy of the network unless the timestep is short enough
that the difference $F_i$ is computed accurately for each population in the
network.  But this restriction means that the maximum timestep is set by the
largest fluxes (that is, the largest stable timestep is determined by the
inverses of the highest rates); this lands us back in the explicit integration
conundrum that the maximum stable timestep is set by the fastest transitions,
even if the primary interest is in quantities varying on a much longer
timescale.

Thus, in the approach to equilibrium the problem with explicit integration is
not negative populations directly, but an unacceptable loss of accuracy that may
occur even before any populations become negative.  This is still a stiffness
issue because it involves stability in systems with disparate timescales. In
this case the contrasting timescales are the very rapid reactions driving the
system to equilibrium relative to the very slow timescale associated with
equilibrium itself. Thus any system nearing equilibrium can be expected to
exhibit this form of stiffness instability. As we now address, this distinction
among sources of stiffness is critical because these stiffness instabilities
have fundamentally different causes and thus fundamentally different solutions.
Furthermore, we shall find that the second class of instabilities can be divided
into two subclasses requiring different stabilizing approximations. The
approximations that we shall introduce in all these cases take care naturally of
the first class of stiffness instabilities because they will prevent the
occurrence of negative probabilities.

\section{\label{sh:approachEquil} Curing Stiffness in the Approach to
Equilibrium}

We shall take equilibration to mean a situation where populations in the network
are being strongly influenced by canceling terms  on the right sides of the
differential equations \eqnoeq{equilDecomposition}.   In terms of the coupled
set of differential equations describing the network, we may distinguish two
qualitatively different conditions:

\begin{enumerate}
 \item 
An equilibration acting at the level of individual differential equations that
we shall call {\em macroscopic equilibration}.
\item
An equilibration acting at the level of subsets of terms within a given
differential equation that we shall term {\em microscopic equilibration}.
\end{enumerate}
Let us consider each of these cases in turn.

\subsection{\label{ss:macroscopic} Macroscopic Equilibration}

The differential equations that we must solve take the general form given in
\eq{equilDecomposition}, $dy_i/dt = \Fplus i - \Fminus i$. One class of
approximate solutions depends upon assuming that $\Fplus i - \Fminus i
\rightarrow 0$ (asymptotic approximations) or $\Fplus i - \Fminus i \rightarrow
$ constant (quasi-steady-state approximations).  We shall term this macroscopic
equilibration, since these conditions involve the entire right side of a
differential equation in \eq{equilDecomposition} tending to zero or a finite
constant. In the first two papers of this series \cite{guidAsy,guidQSS} we
employed asymptotic and quasi-steady-state approximations that removed
entire differential equations from the numerical integration for a network
timestep by replacing them with algebraic approximate solutions for that
timestep. These approximations integrate the full original set of differential
equations, but they reduce the number of equations integrated {\em numerically}.
 This removal of equations from the numerical integration reduces the stiffness 
because it generally decreases the range of timescales in the numerical
integration.

\subsection{\label{ss:microscopic} Microscopic Equilibration}

In \eq{equilDecomposition}, $\Fplus i$ and $\Fminus i$ for a given species $i$
each consist of a number of terms depending on the various populations in the
network, 
\begin{eqnarray}
\deriv{y_i}{t} &=&
\Fplus{i} - \Fminus{i} 
= \sum_j (f^+_j - f^-_j)_i.
\label{equilDecomposition2}
\end{eqnarray}
At the more microscopic level, groups of individual terms on the right side of
\eq{equilDecomposition} in the sum over $j$ may come approximately into
equilibrium (so that the sum of their fluxes is approximately zero), even if the
macroscopic conditions for equilibration are not satisfied and  asymptotic or
quasi-steady-state approximations are not well justified for the species $i$.
The simplest possibility is that forward--reverse reaction pairs such as $A+B
\rightleftharpoons C$, which will contribute flux terms with opposite signs on
the right sides of differential equations in which they participate, come
approximately into equilibrium. As we shall demonstrate, this introduces new
(often fast) timescales into the problem that are at best only partially removed
by asymptotic and quasi-steady-state (QSS) approximations. The new sources
of stiffness associated with these microscopic equilibration effects  explain
why asymptotic and QSS approximations remain accurate but their timestepping
becomes very inefficient in the approach to equilibrium.

As we elaborate in the remainder of this paper, this new source of stiffness
associated with close approach to microscopic equilibration requires a new
algebraic approximation that removes groups of such terms from the numerical
integration by replacing their sum of fluxes with zero. Such an approximation
will not generally reduce the number of equations to be integrated numerically
in a network timestep, but can (dramatically, as we shall see) reduce the
stiffness of those equations by systematically removing terms with fast rates
from the equations.  This reduces the disparity between the fastest and
slowest timescales in the system. Thus shall we convert the approach to
equilibrium in an explicit integration from a liability into an asset. As part
of this elaboration we shall also reach two important general conclusions:
(1)~Approximations based on microscopic equilibration are much more efficient at
removing stiffness than those based on macroscopic equilibration, because they
more precisely target the sources of stiffness in the network. (2)~Macroscopic
and microscopic approximations can complement each other in removing stiffness
from the equations to be integrated numerically and thus the two together are
more powerful than either used alone.

\section{Methods for Partial Equilibrium}
\protect\label{partialEq}

Let us begin to develop some methods to deal effectively with explicit
integration in the approach to equilibrium.  In doing so we draw substantially
on the work of David Mott \cite{mott99}, but we will extend these methods and
obtain much more favorable results for extremely stiff networks than those
obtained in the original work of Mott and collaborators.

The basic idea of partial equilibrium (PE) methods is to inspect the source
terms $\fplus i$ and $\fminus i$ associated with individual reaction pairs in
the network for approach to equilibrium (instead of the composite flux terms
$\Fplus i$ and $\Fminus i$ that are the basis for asymptotic and
quasi-steady-state approximations). Once a fast reaction pair nears equilibrium,
its source terms are removed from the direct numerical integration in the
ordinary differential equations and its effect is incorporated through an
algebraic constraint implied by the equilibrium condition.  Those reactions not
in equilibrium still contribute to the net fluxes for the numerical integrator,
but once the fast reactions associated with an equilibrating reaction pair are
decoupled from the numerical integration the remaining system typically becomes
much less stiff. To illustrate, consider a representative 2-body reaction,
\begin{equation}
 a + b \rightleftharpoons c + d.
\label{partial1.1}
\end{equation}
The source term for this reaction pair can be expressed in the general form
\begin{equation}
 \PEsource{ab}{cd}  = \pm(f_{a+b\rightarrow c+d} - f_{c+d \rightarrow a+b})
= \pm (k\tsub f y_a y_b - k\tsub r y_c
y_d),
\label{partial1.2}
\end{equation}
where the $y_i$ denote population variables for the species $i$ and the $k$s
are rate parameters. This source term will contribute to the right side of the
differential equations describing the change in population for all four species
$a, b, c, d$:
\begin{eqnarray}
 \deriv{y_a}t \mwgalign= \PEsource{ab}{cd} 
+ {\rm other\ terms\ changing\ } y_a
\\
 \deriv{y_b}t \mwgalign= \PEsource{ab}{cd} 
+ {\rm other\ terms\ changing\ } y_b
\\
 \deriv{y_c}t \mwgalign= -\PEsource{ab}{cd} 
+ {\rm other\ terms\ changing\ } y_c
\\
 \deriv{y_d}t \mwgalign= -\PEsource{ab}{cd} 
+ {\rm other\ terms\ changing\ } y_d
\label{partial1.3}
\end{eqnarray}
Thus removing or reducing the stiffness associated with this single reaction
pair influences a whole set of populations and associated reactions, and so can
reduce the overall stiffness of the system. Strictly, the reaction pair of
\eq{partial1.1} is in equilibrium if $\PEsource{ab}{cd} = 0$. Of greater
interest will be  partially-equilibrated systems, where some reactions maintain
$f \sim 0$ as the system evolves but others have $f \ne 0$ (and the members of
these two sets may change over time). Our goal will be to develop methods to
integrate such systems using asymptotic or quasi-steady-state algorithms, but
for modified equations of reduced stiffness in which the flux sums for the
$f\sim 0$ reactions have been replaced by equilibrium constraints.

From Eqs.\ \eqnoeq{partial1.1}--\eqnoeq{partial1.3} the single reaction pair
$a+b \rightleftharpoons c+d$ appears to have four characteristic timescales
associated with the rate of change for the four populations $y_a$, $y_b$, $y_c$,
and $y_d$, respectively. For $a+b \rightleftharpoons c+d$ considered in
isolation, the differential equation governing the abundance of species $a$ is
$$
\deriv{y_a}t = F^+_a - F^-_a = k\tsub r y_c y_d - k\tsub f y_a y_b
= F^+_a - \frac{y_a}{\tau_a},
$$
where $\tau_a \equiv 1/(k\tsub f y_b)$. If we assume the abundances of $c$ and
$d$ and the rate parameters to remain approximately constant in a timestep, the
timescale $\tau_a$ characterizes the rate of change of the species $a$.
Likewise, for the other three  components of the reaction pair we may write
similar differential equations and define similar timescales,
$$
 \tau_a = \frac{1}{k\tsub f y_b} \quad
 \tau_b = \frac{1}{k\tsub f y_a} \quad 
\tau_c = \frac{1}{k\tsub r y_d} \quad
 \tau_d = \frac{1}{k\tsub r y_c}.
$$
Our first task is to construct a {\em single} timescale that characterizes
equilibration of a reaction pair such as $a+b \rightleftharpoons c+d$,
considered in isolation. To do so we introduce the idea of {\em conserved
scalars} \cite{mott99}.

\subsection{Conserved Scalars and Progress Variables}
\protect\label{progvar}

As a simple initial illustration, consider the reaction pair
  $a \rightleftharpoons 2b$,
which has a source term
\begin{equation}
 \PEsource{a}{2b}  = k\tsub f y_a -k\tsub r y_b^2.
\label{progress1.2}
\end{equation}
If no other reactions altered the populations $y_a$ and $y_b$,
\begin{equation}
\deriv{y_a}{t} = -\PEsource{a}{2b}\qquad
\deriv{y_b}{t} = 2\PEsource{a}{2b}.
\label{progress1.2b}
\end{equation}
Thus $2 \diffelement{y_a}/\diffelement t + \diffelement{y_b}/
\diffelement t = 0$ and
$2 y_a + y_b = {\rm constant}$.
The quantity $2 y_a + y_b$ is an example of a conserved scalar.  It is conserved
by virtue of the structure of $a \rightleftharpoons 2b$, not by any particular
dynamical assumptions; thus conservation of this quantity is independent of
whether the reaction is near equilibrium or not. It is convenient to introduce
new variables that are the difference between the initial values of $y_i$ and
their current values
$$
\delta y_a \equiv y_a - y_a^0 \qquad
\delta y_b = y_b - y_b^0.
$$
The initial values $y_a^0$ and $y_b^0$ are constants so the differential
equations \eqnoeq{progress1.2b} can then be written as
\begin{equation}
\deriv{\delta y_a}{t} = -\PEsource{a}{2b}
\qquad
\deriv{\delta y_b}{t} = 2\PEsource{a}{2b}.
\label{progress1.3b}
\end{equation}
This suggests defining a {\em progress variable} $\lambda$ for the
reaction characterized by $\PEsource{a}{2b}$ which satisfies
\begin{equation}
 \deriv{\lambda}{t} = \PEsource{a}{2b} \qquad \lambda_0
\equiv \lambda(t=0) = 0.
\label{progress1.4}
\end{equation}
By comparing \eq{progress1.4} with \eq{progress1.3b} we have
$\lambda = -\delta y_a = \tfrac12 \delta y_b$, so that
\begin{eqnarray}
y_a = \delta y_a + y_a^0 = -\lambda + y_a^0
\qquad
y_b = \delta y_b + y_b^0 = 2\lambda + y_b^0.
\label{progress1.5}
\end{eqnarray}
Thus Eqs.\ 
\eqnoeq{progress1.2} and \eqnoeq{progress1.2b} may be rewritten in terms of
the {\em single variable} $\lambda$ and the solution of two differential
equations for two unknowns is reduced to the solution of \eq{progress1.4} for a
single unknown $\lambda$ [which may then be used to compute $y_a$ and $y_b$
through \eq{progress1.5}].

Let's now apply these ideas to the general 2-body reaction $ a + b
\rightleftharpoons c + d$ of \eq{partial1.1}. Assuming conservation of particle
number, it is clear that the following constraints apply to this reaction
\begin{equation}
 y_a-y_b = c_1 \qquad y_a + y_c = c_2 \qquad y_a + y_d = c_3,
\label{2body1.1}
\end{equation}
where the $c_i$ are constants. Losing one $a$ by this reaction requires the
simultaneous loss of one $b$, so their difference must be constant, and every
loss of one $a$ produces one $c$ and one $d$, which explains the second and
third equations.  These constraints follow entirely from the structure of the
reaction and are independent of dynamics. The constants can be evaluated by
substituting the initial abundances into \eq{2body1.1}, 
$$
c_1 = y_a^0 - y_b^0
\qquad
c_2 = y_a^0 + y_c^0
\qquad
c_3 = y_a^0 + y_d^0.
$$
The differential equation for $y_a$ is
\begin{equation}
 \deriv{y_a}{t} = -k\tsub f y_a y_b + k\tsub r y_c y_d,
\label{2body1.2}
\end{equation}
which can be rewritten using \eq{2body1.1} as 
\begin{equation}
 \deriv{y_a}{t} = a y_a^2 + b y_a + c,
\label{2body1.3}
\end{equation}
where
$$
a = k\tsub r - k\tsub f \qquad b = -k\tsub r(c_2+c_3) + k\tsub f c_1
\qquad c = k\tsub r c_2 c_3.
$$
As we demonstrate below, the approach to equilibrium for any 2-body reaction
pair can be described by a differential equation of this form, and the approach
to equilibrium for any 3-body reaction pair can be approximated by a
differential equation of this form. If $a=0$ the solution of this differential
equation gives the quasi-steady-state (QSS) solution, which we have already
examined in the second paper of this series \cite{guidQSS}. For the general case
$a\ne 0$, let us define a quantity
\begin{equation}
 q \equiv 4ac-b^2\le 0 .
\label{2body1.3a}
\end{equation}
The case $q=0$ is the trivial solution where all abundances are zero, 
so we are interested in solving \eq{2body1.3} for negative values of $q$. 
The general solution for $a\ne0$ and $q < 0$ is \cite{mott99,feg11b}
\begin{equation}
  y_a(t) = -\frac{1}{2a}\left(
b + \sqrt{-q} \,
\frac{1+\phi \exp (-\sqrt{-q} \,t)} {1-\phi \exp (-\sqrt{-q} \, t)},
\right)
\label{2body1.4}
\end{equation}
where
\begin{equation}
 \phi = \frac{2a y_0 + b + \sqrt{-q}} {2a y_0 + b - \sqrt{-q}}.
\label{2body1.5}
\end{equation}
The equilibrium solution corresponds to the limit $t \rightarrow \infty$
of \eq{2body1.4}:
\begin{equation}
 \bar y_a \equiv y^{{\rm\scriptstyle eq}}_a = -\frac{1}{2a} (b + \sqrt{-q}).
\label{2body1.6}
\end{equation}
By analogy with the earlier discussion we define a progress variable
\begin{equation}
 \lambda(t) = -y_a(t) + y_a^0 .
\label{2body1.7}
\end{equation}
Once $y_a$ has been determined the constraints \eqnoeq{2body1.1} may
be used to determine the other abundances:
$$
y_b(t) = y_a(t) -c_1
\quad
y_c(t) = c_2 -y_a(t)
\quad
y_d(t) = c_3 - y_a(t).
$$
The approach of the reaction $a+b \rightleftharpoons c+d$ to equilibrium is thus
governed by a single differential equation \eqnoeq{2body1.3}, which may be
expressed in terms of either a single one of the abundances $y_i$, or the
progress variable $\lambda$ defined in \eq{2body1.7}. The general solution of
this differential equation is given by \eq{2body1.4}, from which the rate at
which the reaction $a+b \rightleftharpoons c+d$ evolves toward the equilibrium
solution \eqnoeq{2body1.6} is determined by a {\em single timescale}
\begin{equation}
 \tau = \frac{1}{\sqrt{-q}}.
\label{2body1.8}
\end{equation}
These equilibrium timescales are illustrated in \fig{equilTimescaleComposite}.
 \putfig
     {equilTimescaleComposite}   
     {0pt}
     {\figdn}
     {0.73}
     {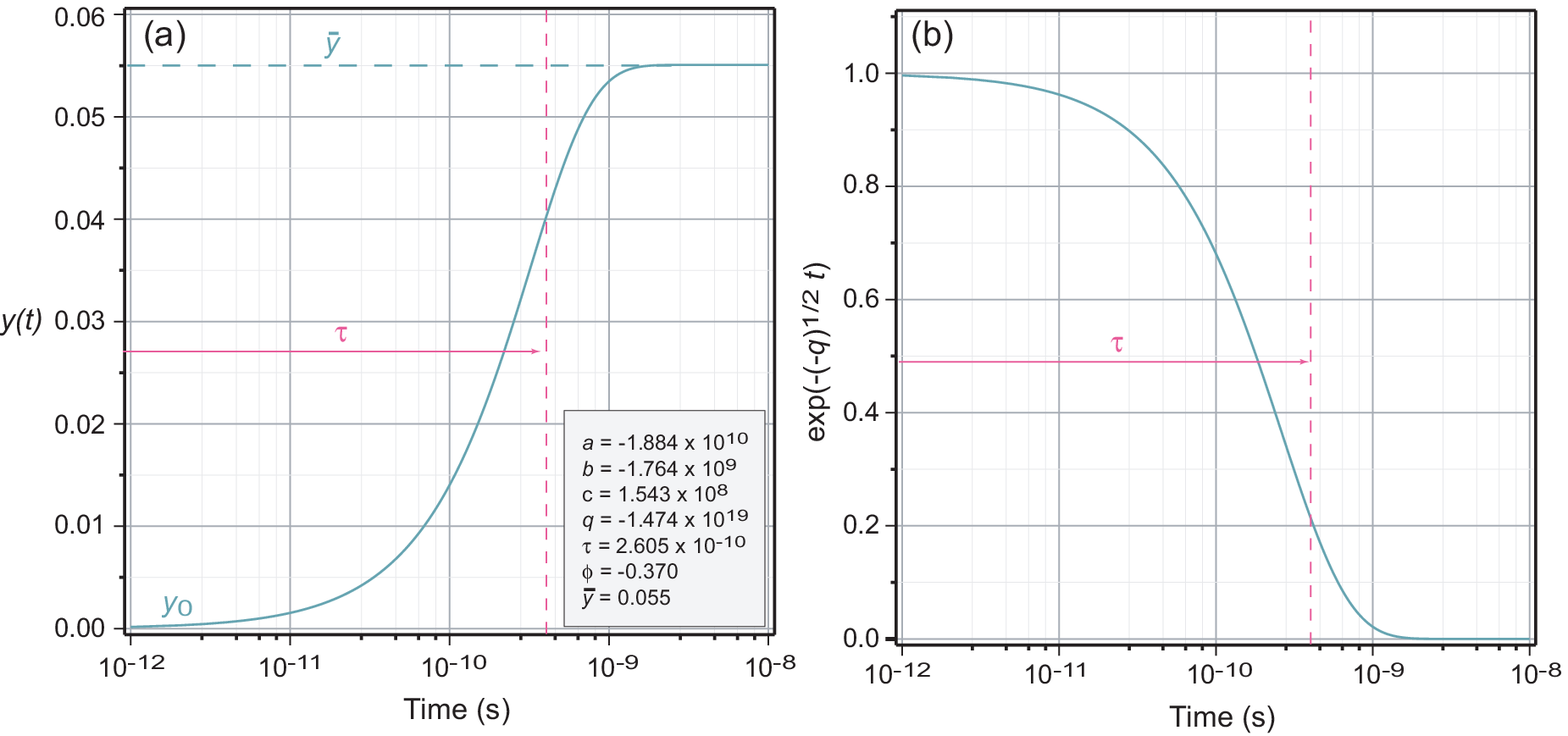}
{(a)~Time evolution of the solution \eqnoeq{2body1.4} assuming $a$, $b$, and $c$
to remain constant. The characteristic timescale for approach to equilibrium
defined by \eq{2body1.8} is labeled $\tau$ and the equilibrium value of $y(t)$
defined by \eq{2body1.6} is denoted by $\bar y$. To illustrate we have assumed
the initial value $y_0 = 0$. For times considerably larger than $\tau$ the
general solution \eqnoeq{2body1.4} saturates at the equilibrium solution
\eqnoeq{2body1.6}. (b)~Behavior of the exponential factor in \eq{2body1.4}.}
We may then estimate whether a reaction is near equilibrium at time $t$ by
requiring
\begin{equation}
 \frac{| y_i(t) - \bar y_i |}{\bar y_i}
< \epsilon_i
\label{2body1.8b}
\end{equation}
for each species $i$ involved in the reaction, where $y_i(t)$ is the actual
abundance, $\bar y_i$ is the equilibrium abundance computed from \eq{2body1.6},
and the user-specified tolerance $\epsilon_i$ can depend on $i$ but will be
taken to be the same for all species in the simplest implementation.
Alternatively, the equilibrium timescale \eqnoeq{2body1.8} in comparison with
the current numerical timestep may be used as a measure of whether a reaction is
near equilibrium.  If $\tau$ is much less than the timestep, it is likely that
equilibrium can be established and maintained in successive timesteps, even if
it is being continually disturbed by other non-equilibrated processes.

\subsection{Reaction Vectors}
\protect\label{reactionvectors}

In a large reaction network there could be thousands of reactions to be examined
for their equilibrium status at each timestep, so implementation of a partial
equilibrium approximation requires a substantial amount of
bookkeeping. Our task will be facilitated by systematic ways to catalog and
examine the equilibrium status of reactions.  We employ a formalism,
adapted from the thesis work of David Mott \cite{mott99}, that 
exploits the analogy of a reaction network to a linear vector space. This will
have two large advantages for us:  (1)~It will place at our disposal
well-established mathematical tools.  (2)~Treating the reaction network as a
linear vector space will permit formulation of a partial equilibrium algorithm
that is not tied closely to the details of a particular problem, thus aiding
portability across disciplines.

It is useful to view the concentration variables for the $n$ species $A_i$ in a
network as components of a composition vector
\begin{equation}
  \bm y = (y_1, y_2, y_3, \ldots y_n),
\label{vector1.1}
\end{equation}
which lies in an $n$-dimensional vector space $\Phi$.  The components $y_i$ can
be any parameters proportional to number densities for the species labeled by
$A_i$, and a specific vector in this space defines a particular composition. 
Any reaction in the network can then be written in the form
\begin{equation}
 \sum_{i=1}^n a_i A_i \rightleftharpoons \sum_{i=1}^n b_i A_i
\label{vector1.2}
\end{equation}
for some sets of coefficients $\{a_i\}$ and $\{b_i\}$.  For example, 
consider the CNO cycle of Fig.~\ref{fig:cnoCycle} and choose
an ordering
\begin{equation}
\bm y = (p \equiv \isotope 1H, \alpha \equiv \isotope{4}{He}, \isotope{12}C,
\isotope{13}N, 
\isotope{13}C, \isotope{14}N,
\isotope{15}O, \isotope{15}N).
\label{cnoVector}
\end{equation}
(Note that usually the positrons $\beta^+$ and the gamma rays $\gamma$ emitted
in \fig{cnoCycle} are not tracked explicitly in the network.) Letting the
composition variables $y_i$ correspond to the mass fraction $X_i$ for a species,
the  $\isotope{12}{C}(p, \gamma)\isotope{13}{N}$ reaction (that is,
$\isotope{1}{H} + \isotope{12}{C} \rightarrow \isotope{13}N$) corresponds to
\eq{vector1.2} with
$
a = \{1, 0, 1, 0, 0, 0, 0, 0 \}
$ and
$b = \{ 0,0,0,1,0,0,0,0\}.
$
The coefficients on the two sides of a reaction may be used to define a vector
$\bm r \in \Phi$ that has the form
$$ 
\bm r = (b_1-a_1, b_2-a_2, \ldots, b_n-a_n),
$$
and defines a composition displacement vector in $\Phi$ associated with the
reaction. For example, for the reaction $\isotope{1}{H} + \isotope{12}{C}
\rightarrow \isotope{13}N$ the components of $\bm r$ are
$
(-1, 0, -1, 1, 0, 0, 0, 0).
$

\subsection{Conservation Laws}
\protect\label{conlaws}

Given an initial composition $\bm y_0 = (y^0_1, y^0_2, y^0_3, \ldots y^0_n)$,
the composition that can be produced by a single pair of reactions labeled by
$i$ is of the form $\bm y = \bm y_0 + \alpha_i \bm r_i$, where $\alpha_i$
is some scalar quantity and for a set of $k$ reactions the final composition
must be of the form
\begin{equation}
 \bm y = \bm y_0 + \sum_{i=1}^k \alpha_i \bm r_i
\label{vector1.4}
\end{equation}
Define a time-independent 
vector $\bm c = (c_1, c_2, c_3, \ldots c_n) \in \Phi$
that is orthogonal to each of the $k$ vectors $\bm r_i$ in
\eq{vector1.4}.  From \eq{vector1.4}
$$
\bm c \cdot (\bm y - \bm y_0) = \bm c \cdot \sum_{i=1}^k \alpha_i \bm r_i
= 0,
$$
which implies that $\bm c \cdot \bm y = \bm c \cdot \bm y_0$ and therefore
that
\begin{equation}
 \sum_{i=1}^n c_i y_i = \sum_{i=1}^n c_i y_0^i = \units{constant}.
\label{vector1.5}
\end{equation}
Thus any vector $\bm c$ orthogonal to the reaction vectors $\bm r_1$, $\bm r_2$,
\ldots $\bm r_k$ gives a linear combination of species abundances that is {\em
invariant under the reactions defined by the vectors $\bm r_i$.} Equation
\eqnoeq{vector1.5} defines a {\em conservation law} following only from the
structure of the network, independent of any dynamical considerations, and thus
must be valid irrespective of dynamical conditions in the network. The conserved
quantities may be determined by forming a matrix $\bm r$ having rows
corresponding to the reaction vectors $\bm r_i$, and solving the matrix equation
$\bm r\cdot \bm c = 0$ for the vector $\bm c$.

\subsection{Example: CNO Cycle}
\protect\label{cnocycle}

Let us illustrate some of the preceding ideas  by
applying the classification scheme
to the network describing evolution of the main part of the CNO cycle with
time. From \fig{cnoCycle}, the reactions of the closed cycle are
\begin{eqnarray}
1:\  p + \isotope{12}{C}\ \rightarrow \ \isotope{13}N + \gamma
&\qquad
2:\  \isotope{13}N \ \rightarrow \ \isotope{13}C 
    + \beta^+ + \nu_e
\nonumber
\\
3:\  p + \isotope{13}C \ \rightarrow \ \isotope{14}N + \gamma
&\qquad
4:\  p + \isotope{14}N \ \rightarrow \ \isotope{15}O + \gamma
\label{vector1.6}
\\
5:\ \isotope{15}O \ \rightarrow \ \isotope{15}N 
    + \beta^+  + \nu_e
&\qquad
6:\  p + \isotope{15}N \ \rightarrow \ \alpha + \isotope{12}C
\nonumber
 \end{eqnarray}
which we shall assume to not be reversible under the temperature and density
conditions characteristic of the CNO cycle. (Since the reactions are assumed to
not be reversible, we shall not apply the partial equilibrium approximation to
this set of reactions. But that is irrelevant for the classification scheme,
which is independent of dynamical conditions in the network.)
Utilizing the vector \eqnoeq{cnoVector}, the source terms for the reactions
\eqnoeq{vector1.6} are
\begin{eqnarray}
 1:\ F_{13} = k_{13} y^1 y^3 
&\qquad
2:\ F_4 = k_4 y^4
&\qquad
3:\ F_{15} = k_{15} y^1 y^5
\nonumber
\\
4:\ F_{16} = k_{16} y^1 y^6
&\qquad
 5:\ F_7 = k_7 y^7
&\qquad
 6:\ F_{18} = k_{18} y^1 y^8.
\nonumber
\label{vector1.8}
\end{eqnarray}
where the indices refer to the positions in the vector \eqnoeq{cnoVector}
and the $k$s are rate parameters that in the general case depend on temperature
and density. With the ordering \eqnoeq{cnoVector} the reaction vectors
corresponding to \eq{vector1.6} are
 \begin{eqnarray}
\bm r_1 = (-1, 0, -1, 1, 0,0,0,0)
&\qquad&
\bm r_2 = (0, 0, 0, -1, 1, 0, 0, 0)
\nonumber
\\
\bm r_3 = (-1, 0, 0, 0, -1, 1, 0, 0)
&\qquad&
\bm r_4 = (-1, 0, 0, 0, 0, -1, 1, 0)
\label{vector1.9}
\\
\bm r_5  = (0,0,0,0,0,0,-1,1)
&\qquad&
\bm r_6  = (-1, 1,1, 0,0,0,0,-1)
\nonumber
 \end{eqnarray}
By forming a matrix $\bm r$ having rows given by the reaction vectors $\bm r_i$
and solving the matrix equation $\bm r \cdot \bm c = 0$ for the vector $\bm c$
by gaussian elimination, we find two useful conservation laws, specified by the
vectors $\bm c_1$ and $\bm c_2$.  The first corresponds to conserving the total
number of neutrons plus protons (conservation of nucleon number).  The second
corresponds to conservation of the sum of number densities for all the carbon,
nitrogen, and oxygen (CNO) isotopes in the network, which represents an elegant
derivation of the well-known property that the CNO isotopes catalyze the CNO
cycle, but are not consumed by it.

\subsection{Reaction Group Classes}
\protect\label{reactionGroupClasses}

It will prove useful to associate inverse reaction pairs in {\em reaction group
classes} ({\em reaction groups,} or {\em RG} for short). We employ the
individual reaction classifications used in the REACLIB library \cite{raus2000}
that are illustrated in Table \ref{tb:reaclibClasses}.  In this classification
reactions are assigned to eight categories, depending on the number of {\em
nuclear species} on the left and right side of the reaction equation.  From this
classification it is clear that there are five independent ways that the
reactions of Table \ref{tb:reaclibClasses} can be combined to give reversible
reaction pairs. This forms the basis for the {\em reaction group classification}
illustrated in Table \ref{tb:reactionGroupClasses}.  For example, reaction group
class B consists of reactions from REACLIB reaction class 2 (a $\rightarrow$ b +
c) paired with inverse reactions (b + c $\rightarrow$ a), which corresponds to
REACLIB reaction class 4.

\begin{table}
\caption{\label{tb:reaclibClasses}Reaction classes in the REACLIB
\cite{raus2000} library}
\begin{indented}
\item[]\begin{tabular}{@{}lll}
\br
Class & Reaction & Description or example\\
\mr
            1 &
            a $\rightarrow$ b &
            $\beta$-decay or e$^-$ capture

        \\ 
            2 &
            a $\rightarrow$ b + c &
            Photodisintegration
                                 + $\alpha$ 
	\\ 
            3 &
            a $\rightarrow$ b + c + d&
            $^{12}$C $\rightarrow$ 3$\alpha$ 
	\\ 
            4 &
            a + b $\rightarrow$ c &
            Capture reactions
	\\ 
            5 &
            a + b $\rightarrow$ c + d &
            Exchange reactions
	\\ 
            6 &
            a + b $\rightarrow$ c + d + e &
            $^2$H + $^7$Be $\rightarrow$ $^1$H + 2$^4$He 
	\\ 
            7 &
            a + b $\rightarrow$ c + d + e + f &
            $^3$He + $^7$Be $\rightarrow$ 2$^1$H 
            + 2$^4$He 
	\\ 
            8 &
            a + b + c $\rightarrow$ d \, (+ e) &
            Effective 3-body reactions  
        \\        
\br       
\end{tabular}
\end{indented}
\end{table}

\begin{table}
\caption{\label{tb:reactionGroupClasses}Reaction group classes}
\begin{indented}
\item[]\begin{tabular}{@{}lll}
\br
Class & Reaction pair & REACLIB class pairing\\
\mr
            A &
            a $\rightleftharpoons$ b &
            1 with 1

        \\ 
            B &
            a + b $\rightleftharpoons$ c &
            2 with 4
	\\ 
            C &
            a +b + c $\rightleftharpoons$ d&
            3 with part of 8
	\\ 
            D &
            a + b $\rightleftharpoons$ c + d &
            5 with 5
	\\ 
            E &
            a + b $\rightleftharpoons$ c + d  + e&
            6 with part of 8
        \\        
\br       
\end{tabular}
\end{indented}
\end{table}

Reaction group class A corresponds mostly to $\beta$-decays and their inverses,
so it is important in partial equilibrium calculations only if neutrino
reactions are included. There are only a few reaction pairs of broad importance
in classes C and E other than triple-$\alpha$ ($3\alpha \rightleftharpoons
\isotope{12}C$), so in many practical applications the most important reaction
groups lie predominantly in reaction group classes B and D.  The reason that
REACLIB class 8 appears in both reaction group classes C and E is the ambiguity
in its definition in Table \ref{tb:reaclibClasses}, since it contains reactions
that can have either one or two nuclear products on the right side.  Notice that
reaction group class D is composed of REACLIB class 5 paired with itself, and
that there are no reaction group classes that involve REACLIB reaction class 7
because it has four nuclear species on the right side and REACLIB contains no
reactions with four bodies on the left side of the reaction equation to pair
with it.

Rare exceptions to these observations can occur if a reaction involves a
catalyst (same species appearing on both sides of the equation with the same
coefficients).  For example, consider the reactions
$
n  + 2\isotope{4}{He} \rightleftharpoons \isotope{9}{Be}
$
and
$
p + \isotope9{Be} \rightarrow n + p +2\isotope4{He}.
$
The first reaction pair is classified as reaction group class C since it pairs a
REACLIB class 8 reaction with a REACLIB class 3 reaction. The components of this
reaction pair have reaction vectors that differ from each other by a sign.  The
$p+\isotope9{Be}$ reaction is REACLIB class 7, which normally does not have a
REACLIB class to pair with in equilibrium.  However,  the net effect of this
reaction on populations is the same as $\isotope9{Be} \rightarrow n
+2\isotope4{He}$ since the proton appearing on both sides of the equation is
catalytic and cancels in the population changes.  Thus $p + \isotope9{Be}
\rightarrow n + p +2\isotope4{He}$  has the same reaction vector as
$\isotope9{Be} \rightarrow n +2\isotope4{He}$ and is effectively also the
inverse of $n  + 2\isotope{4}{He}$.  
The only other such anomalous example contained in REACLIB \cite{raus2000} is
afforded by the reactions
$
n + p \rightleftharpoons d
$
and
$
p + d \rightarrow n + p + p,
$
where $d$ denotes the deuteron $\isotope 2H$.

For each reaction group class the  differential equation governing the reaction
pair  takes the form given by \eq{2body1.3}, $dy/dt = ay^2 + by +c$, where $y$
is either a variable proportional to a number density for one of the reaction
species, or a progress variable that measures the change in initial abundances
associated with the reaction pair, and the coefficients $a$, $b$, and $c$ will
be assumed constant within a single network timestep.  An exception occurs for
reaction group classes C and E, which contain 3-body reactions so that the
general form of the differential equation involves cubic terms
$
dy/dt = \alpha y^3 + \beta y^2 + \gamma y + \epsilon.
$
One could take a similar approach as before, solving this cubic equation for the
partial equilibrium properties for the reaction group. However, these ``3-body''
reactions in astrophysics are typically actually sequential 2-body reactions and
we employ an approximation that in any timestep $y(t)^3 \simeq y^{(0)} y(t)^2$,
where the constant $y^{(0)}$ is the value of $y(t)$ at the beginning of the
timestep.  This reduces the cubic equation to an effective quadratic equation of
the form (\ref{2body1.3}), with $a = \alpha y^{(0)} + \beta$, $b = \gamma$, and
$c=\epsilon$. For the case $a\ne 0$ we have already described the corresponding
general solution, associated timescale, equilibrium abundances, and test for
approach to equilibrium of the reaction in Eqs.\
(\ref{2body1.3})--(\ref{2body1.8b}). Our tests suggest that this is a very good
approximation in typical astrophysical environments and we shall treat all
3-body reactions as effective 2-body reactions.

\subsection{Equilibrium Constraints}
\protect\label{equilConstraints}

If a reaction pair from a specific reaction group class of the form
\eqnoeq{vector1.2} is near equilibrium, there will be a corresponding
equilibrium constraint \cite{mott00}
\begin{equation}
 \prod_{i=1}^n y_i^{(b_i-a_i)} = K,
\label{rgclass1.3}
\end{equation}
 where $K$  is some ratio of rate parameters. For example, consider the reaction
group class E  pair $a+b \rightleftharpoons c+d+e$, with differential equations
for the populations $y_i$
$$
\dot y\tsub a = \dot y\tsub b = -\dot y\tsub c = -\dot y\tsub d
= -\dot y\tsub e = -k\tsub f y_a y_b + k\tsub r y_c y_d y_e.
$$
At equilibrium, the requirement that the forward flux $-k\tsub f y_ay_b$ and
backward flux $k\tsub r y\tsub c y\tsub d y\tsub e$ in the reaction pair sum to
zero implies the constraint
$$
\frac{y_a y_b}{y_c y_d y_e} = \frac{k\tsub r}{k\tsub f} \equiv K,
$$
which is of the form \eqnoeq{rgclass1.3}.

\subsection{Reaction Group Classification}
\protect\label{RGclassification}

 Applying the principles discussed in the preceding paragraphs to the reaction
group classes in Table \ref{tb:reactionGroupClasses} gives the results
summarized for reaction group classes A--E in \ref{RGclassificationApp}. This
gives us a complete classification scheme as a basis for applying a partial
equilibrium approximation to arbitrary astrophysical thermonuclear networks.
However, the  methodology is of broader significance. First, since any reaction
compilation in astrophysics could be reparameterized in the REACLIB format, this
classification scheme provides  a partial equilibrium bookkeeping for any
problem in astrophysics. Second, for any large reaction network in any field, 
the classification techniques illustrated here can be applied to group all
reactions into reaction group classes, and to deduce for each reaction group
class the quantities necessary for applying a partial equilibrium approximation.
All that is required is to formulate the network as a linear algebra problem by
choosing a set of basis vectors corresponding to the species of the network, and
then to define the corresponding reaction vectors within this space.
(Mathematically the choice of the vector space is arbitrary as long as it
provides a faithful mapping of possible species and reactions, but physical
interpretation may be aided by judicious choices in specific fields.)  In
principle this need only be done once for the networks of importance in any
particular discipline.

\section{Example Illustrating the Basic Idea}
\protect\label{basicIdeaExample}

Let us work through a simple example illustrating why partial equilibrium, and
partial equilibrium in conjunction with the explicit asymptotic method, could
greatly reduce the stiffness associated with numerical integration of a set
of coupled differential equations. 

\subsection{Network and Reactions}
\protect\label{4elementAlpha}

For this example we consider the simple network
$
 a \rightleftharpoons b \rightleftharpoons c \rightleftharpoons d,
$
including the reaction groups
\begin{equation}
 a + b \rightleftharpoons c
\qquad
3a \rightleftharpoons b
\qquad
a+c \rightleftharpoons d.
\label{simpe1.2}
\end{equation}
Our general conclusions will not be specific to astrophysics, but in fact these
equations have the form of the first part of an astrophysical thermonuclear
alpha-particle network 
$$\alpha
\rightleftharpoons \isotope{12}{C} \rightleftharpoons \isotope{16}{O}
\rightleftharpoons \isotope{20}{Ne},
$$
with $\alpha$-capture ( $\alpha+b \rightarrow c$),
photodisintegration ($c \rightarrow \alpha+b$), and the
triple-$\alpha$ reaction ($3\alpha\rightleftharpoons \isotope{12}{C}$)
included. From the reaction group classification in 
\ref{RGclassificationApp}, these reaction groups belong to classes B and C, and
the associated equilibrium constraints are given in Table
\ref{tb:constraintsForExample},%
\begin{table}
\caption{\label{tb:constraintsForExample}Reaction groups and constraints}
\begin{indented}
\item[]\begin{tabular}{@{}llll}
\br
Group & Reactions & Class & Constraint\\
\mr
            1 &
            $a+b \rightleftharpoons c$ &
            B &
            $y_a y_b = \frac{\kback1}{\kforward1}\,y_c$
        \\  
            2 &
            $3a \rightleftharpoons b$ &
            C &
            $y_a^3 = \frac{\kback2}{\kforward2}\,y_b$
        \\    
            3 &
            $a+c\rightleftharpoons d$ &
            B &
            $y_a y_d = \frac{\kback3}{\kforward3}\,y_d$
        \\   
\br       
\end{tabular}
\end{indented}
\end{table}
where the $k$ are the rate parameters (generally time-dependent), with the
superscripts denoting the reaction group and the subscripts indicating forward
(f) and reverse (r) directions in \eq{simpe1.2}. Thus $k_{\rm\scriptstyle
r}^{(1)}$ is the rate for $c \rightarrow a+b$.
The coupled set of ordinary differential equations to be solved is
\begin{eqnarray}
 \deriv{y_a}{t} \mwgalign= -\kforward{2} y_a^3 + \kback2 y_b
-\kforward1 y_ay_b + \kback1 y_c
- \kforward3 y_a y_c + \kback3 y_d
\label{simpe1.4a}
\\
\deriv{y_b}{t} \mwgalign= \kforward2 y_a^3 - \kback2 y_b + \kback1 y_c -
\kforward1 y_a y_b
\label{simpe1.4b}
\\
\deriv{y_c}{t} \mwgalign= \kforward1 y_a y_b - \kback1 y_c
-\kforward3 y_a y_c +\kback3 y_d 
\label{simpe1.4c}
\\
\deriv{y_d}{t} \mwgalign= \kforward3 y_a y_c - \kback3 y_d.
\label{simpe1.4d}
\label{simpe1.4}
\end{eqnarray}
From Table \ref{tb:constraintsForExample} there are potentially three
constraints available if the system comes into full equilibrium, and one or two
constraints if it is in partial equilibrium.  In addition to the equilibrium
constraints, we may wish to impose constraints associated with conservation laws
for the system, such as preservation of particle number.  In actual applications
this will be very important, but in this example we ignore conservation
laws and concentrate on understanding the role of equilibrium constraints. There
are two general approaches that we might take to using equilibrium constraints
to simplify the solution of the differential equations in Eqs.\
\eqnoeq{simpe1.4a}--\eqnoeq{simpe1.4d}:
\begin{enumerate}
 \item 
Use the constraints to reduce the number of equations that we have to solve in
a given timestep.
\item
Use the constraints to reduce the stiffness of the equations that we have to
solve in a given timestep.
\end{enumerate}
The first approach reduces the number of equations to solve, but in most cases
does  not change the stiffness much for the equations that are solved until
substantial numbers of equations have been removed from the numerical
integration. In the second approach, we still solve the same number of
equations, but the equations that we solve are less stiff than the original ones
because we have modified the structure on the right side of the equalities in
Eqs.\ \eqnoeq{simpe1.4a}--\eqnoeq{simpe1.4d}. We shall see that the second
approach tends to naturally remove the fastest timescales that remain in the
system when it is applied, so it can have a dramatic effect on the stiffness of
the system. Thus in this paper we shall only outline using constraints to reduce
the number of equations, and then concentrate on how constraints can be used to
reduce stiffness.

\subsection{Using Equilibrium Constraints to Reduce the Number of Equations}

Consider the constraint from $a+b \rightleftharpoons c$ written in the form
$
y_a y_b - K y_c =0,
$
where $K_1 \equiv \kback 1/\kforward 1$. Taking the derivative with respect to
time of this expression gives
$$
y_a \deriv{y_b}{t} + y_b \deriv{y_a}{t} - K_1 \deriv{y_c}{t} - y_c
\deriv{K_1}{t} = 0, 
$$
which is a constraint that could be used to eliminate one of Eqs.\
\eqnoeq{simpe1.4a}--\eqnoeq{simpe1.4d} from the numerical integration, say
\eq{simpe1.4b}. Notice that removing \eq{simpe1.4b} leaves all of the rate
parameters of the original problem present in the remaining equations, so it
presumably has had only a small impact on the stiffness.  In a similar manner,
as other reaction pairs come into equilibrium the associated constraints can be
used to remove additional equations from the numerical integration.  We shall
not use this approach in the present context, but we note in passing that this
represents a systematic way to introduce a partial equilibrium approximation
into an implicit-method calculation.  In that case, reducing the number of
equations to be integrated can be significant because it reduces the sizes of
the matrices that must be inverted at each integration step.

\subsection{Using Equilibrium Constraints to Reduce Stiffness}

Instead of using the equilibrium constraints to reduce the number of equations,
let us now outline how to use them to reduce the stiffness of the original set
of equations by applying the constraint directly to the abundances rather than
their derivatives.  Suppose that the reaction $a+b \rightleftharpoons c$ comes
into equilibrium, implying the constraint $y_a y_b = \kback1 y_c /\kforward1.$
Substituting this for $y_a y_b$ in Eqs.\ \eqnoeq{simpe1.4a}--\eqnoeq{simpe1.4d},
various terms cancel and we obtain the modified equations
\begin{eqnarray}
 \deriv{y_a}{t} \mwgalign= -\kforward{2} y_a^3 + \kback2 y_b
- \kforward3 y_a y_c + \kback3 y_d
\label{simpe1.5a}
\\
\deriv{y_b}{t} \mwgalign= \kforward2 y_a^3 - \kback2 y_b
\label{simpe1.5b}
\\
\deriv{y_c}{t} \mwgalign= 
-\kforward3 y_a y_c +\kback3 y_d 
\label{simpe1.5c}
\\
\deriv{y_d}{t} \mwgalign= \kforward3 y_a y_c - \kback3 y_d.
\label{simpe1.5d}
\label{simpe1.5}
\end{eqnarray}
This is the same number of equations as in
\eqnoeq{simpe1.4a}--\eqnoeq{simpe1.4d}, but now $\kforward1$
and $\kback1$ no longer appear on the right sides of the differential equations.
Furthermore, we may expect that the rate parameters that have been removed were
fast ones, because they were responsible for bringing the first reaction group
into equilibrium in the network. Therefore, on general grounds we may expect
that the differential equations Eqs.\ \eqnoeq{simpe1.5a}--\eqnoeq{simpe1.5d} are
less stiff than the original equations \eqnoeq{simpe1.4a}--\eqnoeq{simpe1.4d},
because there will be less disparity between the fastest and slowest rates in
the network.

Let us suppose further that when the fluxes on the right side of Eqs.\
\eqnoeq{simpe1.5a}--\eqnoeq{simpe1.5d} are computed we find that two of the
differential equations---let's say \eqnoeq{simpe1.5a} and
\eqnoeq{simpe1.5b}---satisfy the asymptotic condition. Then these two entire
equations would be solved for the timestep by the algebraic asymptotic formulas,
which eliminates another whole set of terms involving different rate parameters
from the the numerical integration.  Thus we can see conceptually how
simultaneous use of partial equilibrium and asymptotic approximations could
reduce considerably the effective stiffness of a complex set of equations.
Continuing our example, suppose that the reactions $a+b\rightleftharpoons c$
remain in equilibrium (if a reaction group in equilibrium drops out of
equilibrium at a later time, the corresponding flux terms must be restored to
the differential equations) and at a later time the reaction $3a
\rightleftharpoons b$ also comes into equilibrium. Now there are two constraints
to be applied to Eqs.\ \eqnoeq{simpe1.4a}--\eqnoeq{simpe1.4d}, or one new one to
be applied to Eqs.\ \eqnoeq{simpe1.5a}--\eqnoeq{simpe1.5d}. Applying the new
equilibrium constraint $y_a^3 = \kback2 y_b /\kforward2$
to Eqs.\ \eqnoeq{simpe1.5a}--\eqnoeq{simpe1.5d}, we obtain
\begin{eqnarray}
 \deriv{y_a}{t} \mwgalign= - \kforward3 y_a y_c + \kback3 y_d 
\label{simpe1.7a}
\\
\deriv{y_b}{t} \mwgalign=  0
\label{simpe1.7b}
\\
\deriv{y_c}{t} \mwgalign= 
-\kforward3 y_a y_c +\kback3 y_d
\label{simpe1.7c}
\\
\deriv{y_d}{t} \mwgalign= \kforward3 y_a y_c - \kback3 y_d.
\label{simpe1.7d}
\label{simpe1.7}
\end{eqnarray}
Again we have the same number of equations to integrate numerically as
originally (although \eq{simpe1.7b} has become trivial), but additional fast
terms have been removed from the right side of the differential equations,
reducing their stiffness even further. As before, if  any equations in the
preceding set satisfy the asymptotic condition, these equations may be removed
from the numerical integration in favor of an algebraic solution, potentially
further reducing the stiffness of the numerical system.

Finally, suppose that $a+b \rightleftharpoons c$ and $3a \rightleftharpoons b$
remain in equilibrium, and at some later time $a+c \rightleftharpoons d$ comes
into equilibrium too, implying a new constraint $y_a y_c = \kback3 y_d
/\kforward3$. Inserting this expression for $y_a y_b$ into Eqs.\
\eqnoeq{simpe1.7a}--\eqnoeq{simpe1.7d}, we obtain the system characteristic of
complete equilibrium,
\begin{equation}
 \deriv{y_a}{t} = \deriv{y_b}{t} = \deriv{y_c}{t} = \deriv{y_d}{t} = 0.
\label{simpe1.9}
\end{equation}
Of course this set of equations can be integrated trivially, but that is the
point! The systematic application of partial equilibrium techniques to an
intrinsically highly-stiff system has produced an approximately equivalent
system in which all numerical stiffness has been removed (since no timescales 
remain in the equations).  Thus there is no stability restriction on the
timestep, should we choose to integrate \eq{simpe1.9} numerically.

In realistic large and stiff networks we will generally be integrating
numerically in regimes where at least some reactions are not fully in
equilibrium, so the trivial limit of \eq{simpe1.9} will seldom be reached except
for systems very near overall equilibrium. Nevertheless, the preceding example
illustrates that large reductions in stiffness may still be realized through
systematic application of the constraints for those reactions that do come into
equilibrium.  In realistic networks we may expect that in the approach to
equilibrium, situations where the numerical system being integrating is only
somewhat removed from a trivial one of the form given by \eq{simpe1.9} can occur
frequently. In those cases, we may expect large increases in the maximum
explicit integration timestep from systematic application of the constraints
implied by reactions that come into equilibrium, coupled with asymptotic or QSS
approximations employed when conditions warrant it, to the resulting simplified
equations.

\section{General Methods for Partial Equilibrium Calculations}
\protect\label{generalMethods}

We now have a set of tools to implement partial equilibrium
approximations, but there are a number of practical issues that require
resolution before we can make realistic calculations. To that end, let us now
outline a specific approach to applying partial equilibrium methods.

\subsection{Overview of Approach}
\protect\label{approachOverview}

The partial equilibrium method will  be used in conjunction with the
asymptotic approximation in the following examples. (Although we shall not
address it in this paper, a similar algorithm could be employed that replaced
the asymptotic approximation with the quasi-steady-state approximation described
in Ref.\ \cite{guidQSS}.) That is, we shall remove fluxes from
the numerical integration corresponding to reaction pairs that are in
equilibrium, but the integration is still performed formally over the full
reaction network, subject to an asymptotic approximation. Once the reactions of
the network are classified into reaction groups, 
the algorithm has three basic steps: 
\begin{enumerate}
 \item 
During a numerical integration step one begins with the full network of
differential equations, but in computing the net fluxes all terms involving
reaction groups that are judged to be equilibrated (based on criteria determined
by isotopic populations at the end of the previous timestep) are assumed to sum
identically to zero net flux and are omitted from the flux summations.
\item
A timestep $\Delta t$ is then chosen (using a variant of the timestepping
algorithm described in Refs.\ \cite{guidAsy,guidQSS}), and this is used in
conjunction with the fluxes to determine how many isotopes in the network
satisfy the asymptotic condition according to the criteria of Ref.\
\cite{guidAsy}. For those isotopes that are not asymptotic, the change in
abundance for the timestep is then computed by ordinary forward (explicit)
finite difference, but for those isotopes judged to be asymptotic the abundance
changes for the timestep are instead computed using the analytical asymptotic
approximation of Ref.\ \cite{guidAsy}.
\item
Finally, for all isotopes in reaction groups taken to be in equilibrium at the
beginning of the current timestep, it is assumed that reactions not in
equilibrium will have driven these populations slightly away from their
equilibrium values during the numerical timestep.  These populations are then
adjusted, subject to a condition that the sum of the mass fractions remain equal
to one, to restore their equilibrium values at the end of the timestep.
\end{enumerate}
Hence the partial equilibrium  approximation does not reduce 
the number of equations to be integrated but instead removes  the
stiffest parts  of their fluxes in each timestep.  In contrast, the asymptotic
approximation reduces the number of differential equations
integrated numerically within a timestep by replacing the numerical forward
difference with an analytically-computed  abundance  for those
isotopes
satisfying the asymptotic condition.  

\subsubsection{Reducing Stiffness}
\protect\label{reduceStiff}

Both of these approximations reduce stiffness in the  integration by replacing
numerical finite difference with algebraic conditions, but by different and
potentially complementary means.  The partial equilibrium method operates {\em
microscopically} to remove individual components of reaction fluxes that become
stiff in the approach to equilibrium; the asymptotic approximation operates {\em
macroscopically} to remove the entire net flux altering the abundance of
individual isotopes from the numerical integration. The partial equilibrium and
asymptotic approaches are potentially complementary because partial equilibrium
can make the right side of the differential equation for a given isotope  less
stiff, even if the isotope does not satisfy the asymptotic condition, while
conversely the asymptotic condition (by removing the entire flux of selected
differential equations from the numerical update) can effectively remove stiff
reaction components even if they do not satisfy partial equilibrium conditions.
For compactness we shall often refer simply to the partial equilibrium (PE)
approximation in what follows, but it should be understood that this  means
partial equilibrium approximation plus asymptotic approximation as described
above.

\subsubsection{Operator-Split Restoration of Equilibrium}
\protect\label{restoreEquilComments}

The third  step (restoration of equilibrium) in the above algorithm may be
viewed as an operator-split separation of physical timescales {\em within a
single numerical network timestep} (with evolution of the network itself already
being treated on a different level as operator-split from the evolution of the
hydrodynamics). The idea is illustrated schematically in
\fig{operatorSplitEquil}.%
 \putfig
     {operatorSplitEquil}    
     {0pt}
     {\figdn}
     {0.52}
{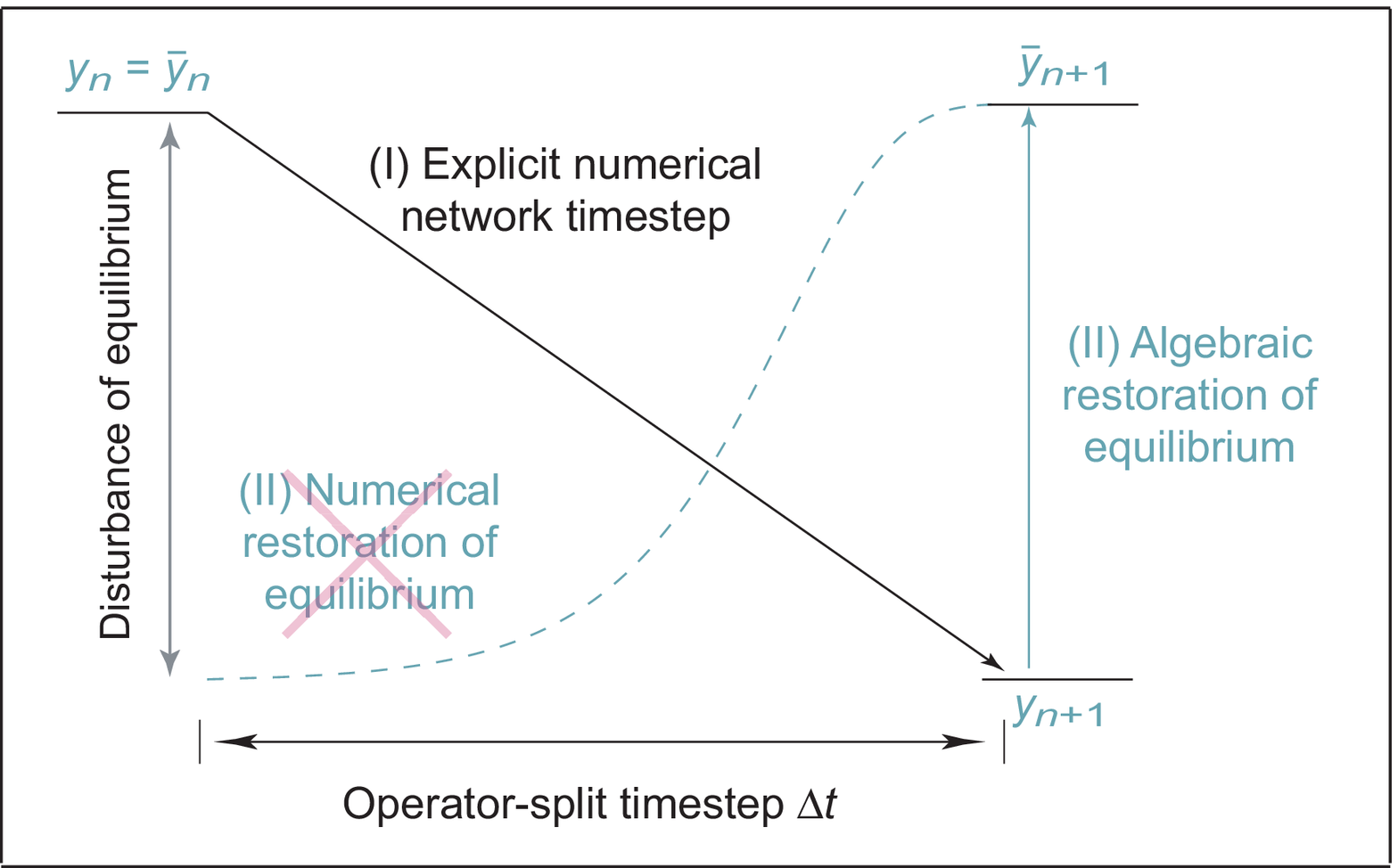}
{Illustration of the ``operator-split'' restoration of equilibrium abundance for
an isotope. At the beginning of a network timestep that advances from $t_n$ to
$t_{n+1}$ the abundance $y_n$ is assumed to take its equilibrium value $\bar
y_n$. An explicit numerical timestep $\Delta t = t_{n+1}-t_n$ is then taken for
all abundances, but with the contribution of equilibrated reaction groups
removed from the fluxes. The isotope in question starts the timestep with its
equilibrium abundance, but the numerical timestep will generally leave the
abundance at a value $y_{n+1}$ that is not equal to the  new equilibrium value
$\bar y_{n+1}$ calculated from \eq{2body1.6} because of its coupling to
non-equilibrium reactions (the effect of the diagonal arrow). If the equilibrium
approximation is to be maintained for the next timestep, the equilibrium value
$\bar y_{n+1}$ must be restored before that timestep is taken. We could in
principle restore the equilibrium by integrating numerically in a second step
(the dashed blue line).  However, we already know the result of that
integration: it is given by \eq{2body1.6}, because by hypothesis $\Delta t$ is
larger than the characteristic equilibration time for all reactions assumed to
be in equilibrium. Thus we accomplish the second operator-split step {\em
algebraically rather than numerically} by replacing the numerically-computed
$y_{n+1}$ with $\bar y_{n+1}$ specified through \eq{2body1.6} (the vertical
arrow on the right side). }
Establishment of equilibrium for individual reaction groups assumed to be in
equilibrium within a single numerical timestep is fast compared with the
timescale for reactions causing net changes in the abundances within a timestep.
Thus, if we were to do the integration exactly, equilibrium for those reaction
groups would be maintained during the timestep; but then we would have a
highly-stiff numerical system combining fast and slow components. Our
approximation replaces this actual evolution by a two-step process: 
\begin{enumerate}
 \item 
Evolve all network abundances by explicit forward differencing and asymptotic
approximations, with the equilibrium reactions assumed to maintain their
equilibrium (net zero) fluxes over the timestep. (This is what justifies
removing the fluxes that are assumed equilibrated from the flux summation.)
\item
Then, at the end of this first step, evolve all abundances assumed to be
participating in equilibrium (which will have been disturbed by non-equilibrium
processes in the first step) from their values computed at the end of the first
step back to equilibrium values using \eq{2body1.6}, while holding
non-equilibrium abundances constant.
\end{enumerate}
Unlike the case for the usual operator-splitting between evolution of the
hydrodynamics and evolution of the reaction network, this second step is {\em
not computed numerically by finite difference} but is instead implemented
through  {\em algebraic constraints.} This is possible because the equilibrium
assumption for a given reaction group means that it should evolve to the
equilibrium solution before the end of the timestep.  Therefore, since we know
how the story ends, rather than integrating it explicitly we may simply replace
the population at the end of the numerical timestep by its equilibrium value,
calculated from \eq{2body1.6}.
We gain from this separation of timescales and solution of the fast
timescale by algebraic constraint rather than a finite-difference approximation
a potentially large reduction of stiffness for the remaining part of the
system that is integrated numerically. We shall demonstrate below that this can
increase by orders of magnitude the maximum stable and accurate
timestep for the overall integration in near-equilibrium conditions.

\subsubsection{Complications in Realistic Networks}
\protect\label{complexityRealNet}

The complication for this basic idea in a realistic network with more than one
reaction group in equilibrium is that there  may be more than one computed
equilibrium value for a given species that is assumed to be in equilibrium. This
is because the equilibrium abundance  \eq{2body1.6} must be computed
separately for each reaction group, and an isotope will generally be found
in more than one reaction group. For example, assume an $\alpha$ network in
which both reaction group I, defined by $\alpha + \isotope{28}{Si}
\rightleftharpoons \isotope{32}{S}$, and reaction group II, defined by $\alpha +
\isotope{40}{Ca} \rightleftharpoons \isotope{44}{Ti}$, are assumed to be in
equilibrium. Then by our algorithm $Y_\alpha$ should assume its equilibrium
value at the end of each timestep, but for each timestep there will be {\em two
values} for the equilibrium abundance $\bar Y_\alpha$, corresponding to
\eq{2body1.6} solved for reaction group I and reaction group II, respectively.
In the general case these values could be different, since the rates entering
into \eq{2body1.6} are different for the two reaction groups, and equilibrium
within each reaction group is specified only to the tolerance implied by
$\epsilon_i$ in \eq{2body1.8b}. More realistically in larger networks a given
isotope may be found in many reaction groups, some in equilibrium and some not,
and for each group in equilibrium there will be a separate computed equilibrium
value for the isotope. 

However, by hypothesis the equilibrium abundances of a given isotope computed
for each of its equilibrated reaction groups cannot be {\em too} different,
since this would violate the equilibrium assumption. For example, consider the
$\alpha$-network example from above. There is only {\em one} actual abundance
$Y_\alpha$ in the network at a given time, and its value must be such that it
satisfies {\em simultaneously} the equilibrium conditions for reaction group I
and reaction group II, within the tolerances of \eq{2body1.8b}; otherwise the
equilibrium assumption would be invalidated. Therefore, restoration of
equilibrium for a given isotope will correspond to setting its abundance to a
compromise choice among each of the (similar) predicted equilibrium values for
all equilibrated reaction groups in which it participates. This is a
self-consistent approximation as long as the spread in possible equilibrium
abundances remains consistent with the tolerances used to impose equilibrium in
\eq{2body1.8b}, for if this were not true it would suggest that the equilibrium
assumption itself is suspect.  

It is  important conceptually that the use of ``equilibrium assumption'' as
convenient shorthand in this discussion not be misinterpreted.  The decision
that a reaction group is in equilibrium is not an arbitrary choice but rather is
made by the full network itself in each timestep.  A reaction group is in
equilibrium if it satisfies \eq{2body1.8b} for every species in the group, so
the only arbitrary choices are the tolerances $\epsilon_i$.  If the condition
\eqnoeq{2body1.8b} is satisfied in a timestep for some reaction group, it is a
statement by the network that in its current state the equilibration timescale
for the reaction group has been found to be fast enough to maintain approximate
equilibrium over the timestep.

\subsection{Specific Methods for Restoring Equilibrium}

As noted above, in realistic networks we must have a systematic way to restore
equilibrium at the end of numerical timesteps since those reaction in
equilibrium at the beginning of the timestep will generally be disturbed
slightly  away from equilibrium during the timestep by those reactions that are
not in equilibrium.  This effect is tiny for one reaction timestep, but will
accumulate to unacceptable error over many timesteps if we do not correct for it
in each timestep.  In this section we outline three potential approaches to this
problem. Although these approaches are general, it probably is easiest to
understand them in the context of a specific application so we shall illustrate
with a 4-isotope alpha network having species $\alpha$, \isotope{12}{C},
\isotope{16}{O}, and \isotope{20}{Ne}, that we already illustrated
schematically in \S\ref{4elementAlpha}.  Using explicit notation for
the alpha network we include the reactions
\begin{equation}
\begin{array}{ll}
 0: &\ 3\alpha \rightleftharpoons \isotope{12}{C}
\qquad
1: \ \alpha + \isotope{12}{C} \rightleftharpoons \isotope{16}{O}
\\[2pt]
2: &\ \isotope{12}C + \isotope{12}C \rightleftharpoons \alpha +
\isotope{20}{Ne}
\qquad
3: \ \alpha + \isotope{16}{O} \rightleftharpoons \isotope{20}{Ne} ,
\end{array}
\label{4alpha1.0}
\end{equation}
which we shall reference below in terms of the number on the left side for each
reaction, and the
differential equations \eqnoeq{simpe1.4a}--\eqnoeq{simpe1.4d} governing the
abundances then become
\begin{eqnarray}
  \dot Y_{\alpha} \mwgalign= -\kforward 0 Y_\alpha^3
+ \kback 0 Y_{12} - \kforward 1 Y_\alpha Y_{12}  + \kback 1 Y_{16}
\nonumber\\ 
\flush  + \kforward 2 Y_{12}^2 -\kback 2 Y_\alpha Y_{20}
- \kforward 3 Y_\alpha Y_{16} + \kback 3 Y_{20}
\label{4alpha1.1a}
\\
\dot Y_{12} \mwgalign= \kforward 0 Y_\alpha^3 - \kback 0 Y_{12}
- \kforward 1 Y_\alpha Y_{12} +\kback 1 Y_{16} - \kforward 2 Y_{12}^2
+\kback2 Y_\alpha Y_{20}
\label{4alpha1.1b}
\\
\dot Y_{16} \mwgalign= \kback1 Y_\alpha Y_{12} -\kback1 Y_{16}
-\kforward3 Y_\alpha Y_{16} + \kback3 Y_{20}
\label{4alpha1.1c}
\\
\dot Y_{20} \mwgalign= \kforward 2 Y_{12}^2 -\kback2 Y_\alpha Y_{20}
+ \kback 3 Y_\alpha Y_{16}
-\kback3 Y_{20},
\label{4alpha1.1d}
\end{eqnarray}
where we use a notation $Y(\isotope{12}C) \equiv Y_{12}$ and so on, for the
abundances (defined in \eq{5.35}), and the rate parameters $\kforward n $ and
$\kback n$ refer to 
forward and reverse rates, respectively, for the reactions pairs labeled by the
numbers on the left sides in \eq{4alpha1.0}. Thus $\kforward 0$ is the rate
parameter for $3\alpha \rightarrow \isotope{12}C$ and $\kback 3$ is the rate
parameter for $\isotope{20}{Ne} \rightarrow \alpha + \isotope{16}{O}$.

\subsubsection{Using Iteration to Reimpose Equilibrium Abundance Ratios}
\protect\label{abundRatioIteration}

If we now impose equilibrium on the reaction $\alpha + \isotope{16}O
\rightleftharpoons \isotope{20}{Ne}$ we obtain the constraint
\begin{equation}
 Y_\alpha Y_{16} = \frac{\kback 3}{\kforward 3} Y_{20} \equiv K Y_{20}.
\label{4alpha1.2}
\end{equation}
Inserting this constraint  into Eqs.~(\ref{4alpha1.1a})--(\ref{4alpha1.1d})
eliminates the last two terms in each of Eqs.~(\ref{4alpha1.1a}),
(\ref{4alpha1.1c}), and (\ref{4alpha1.1d}), giving the reduced equations
\begin{eqnarray}
  \dot Y_{\alpha} \mwgalign= -\kforward 0 Y_\alpha^3
+ \kback 0 Y_{12}  
 - \kforward 1 Y_\alpha Y_{12} + \kback 1 Y_{16} 
+ \kforward 2 Y_{12}^2 -\kback 2 Y_\alpha Y_{20}
\label{4alpha1.3a}
\\
\dot Y_{12} \mwgalign= \kforward 0 Y_\alpha^3 - \kback 0 Y_{12}
- \kforward 1 Y_\alpha Y_{12} +\kback 1 Y_{16} 
- \kforward 2 Y_{12}^2 +\kback2 Y_\alpha Y_{20}
\label{4alpha1.3b}
\\
\dot Y_{16} \mwgalign= \kback1 Y_\alpha Y_{12} -\kback1 Y_{16}
\label{4alpha1.3c}
\\
\dot Y_{20} \mwgalign= \kforward 2 Y_{12}^2 -\kback2 Y_\alpha Y_{20}
\label{4alpha1.3d}
\end{eqnarray}
This is the same number of equations as before, but these equations should now
be less stiff than the original Eqs.~(\ref{4alpha1.1a})--(\ref{4alpha1.1d})
because the terms eliminated by substituting the equilibrium condition
(\ref{4alpha1.2}) are associated with the reaction pair that was first to come
into equilibrium, precisely because it involves fast reactions. 
However, if we impose the equilibrium condition (\ref{4alpha1.2}) at the
beginning of a timestep, it  generally will no longer be satisfied at the end of
the timestep because the other reactions are not in equilibrium and will change
the abundances of $\alpha$, \isotope{16}O, and \isotope{20}{Ne} through
non-equilibrium transitions.  Thus, we must adjust the populations at the end of
the timestep to reimpose the condition (\ref{4alpha1.2}), assuming the
equilibrium condition to still be satisfied for that reaction pair. In addition
to the algebraic constraint (\ref{4alpha1.2}),
we have the condition
\begin{equation}
\Sigma_X \equiv 
 \sum_i X_i = 
4Y_\alpha + 12 Y_{12} + 16Y_{16} + 20Y_{20} = 1,
\label{4alpha1.4}
\end{equation}
imposed by the requirement that the sum of the mass fractions be unity
(conservation of nucleon number; see \eq{5.35}).  In contrast to the constraint
(\ref{4alpha1.2}), which is valid only if partial equilibrium conditions for the
reaction $\alpha + \isotope{16}O \rightleftharpoons \isotope{20}{Ne}$ are
satisfied, the constraint (\ref{4alpha1.4}) is a conservation law that must
always be satisfied. Thus, to reimpose equilibrium we  have available two
conditions, Eqs.~(\ref{4alpha1.2}) and (\ref{4alpha1.4}).

We assume equilibrium at the beginning of the timestep in the $\alpha +
\isotope{16}O \rightleftharpoons \isotope{20}{Ne}$ reaction (only); thus we
advance the solution through a timestep $\Delta t$ by solving the reduced
equations (\ref{4alpha1.3a})--(\ref{4alpha1.3d}) using the explicit asymptotic
algorithm \cite{guidAsy}. Let us denote the (numerically computed) abundances at
the end of this timestep by $\tilde Y_i = (\tilde Y_\alpha, \tilde Y_{12},
\tilde Y_{16}, \tilde Y_{20})$.  At the beginning of the timestep the condition
(\ref{4alpha1.2}) is satisfied, by hypothesis, but at the end of the timestep in
general 
$
 \tilde Y_\alpha \tilde Y_{16} \ne  K \tilde Y_{20}.
$
But the assumption that  $\alpha + \isotope{16}O
\rightleftharpoons \isotope{20}{Ne}$ is in partial equilibrium implies (loosely)
that the characteristic timescale $\tau$ for this reaction group must be less
than the integration timestep.  Thus, we may assume that if we had taken short
enough timesteps the reaction pair $\alpha + \isotope{16}O \rightleftharpoons
\isotope{20}{Ne}$ would have brought the populations for $Y_\alpha$, $Y_{16}$,
and $Y_{20}$ back into equilibrium by the end of the actual timestep taken, and
we use the algebraic conditions given by Eqs.~(\ref{4alpha1.2}) and
(\ref{4alpha1.4}) to reimpose $\alpha + \isotope{16}O \rightleftharpoons
\isotope{20}{Ne}$ equilibrium at the end of the timestep.
Eqs.~\eqnoeq{4alpha1.2} and \eqnoeq{4alpha1.4} allow us to write
\begin{equation}
\begin{array}{l}
F_0 \equiv Y_\alpha -\frac{K}{Y_{16}} Y_{20} = 0
\qquad
F_1 \equiv Y_{16} - \frac{K}{Y_{\alpha}}Y_{20} = 0
\\
F_2 \equiv Y_\alpha + 3\tilde Y_{12} + 4 Y_{16} + 5Y_{20}-\tfrac14 = 0
\end{array}
\end{equation}
This may be written as the vector equation ${\bm F} = 0$, which we can
solve by Newton--Raphson iteration for the unknowns $Y_\alpha$, $Y_{16}$, and
$Y_{20}$, starting the iteration from computed values
$\tilde Y_\alpha$, $\tilde Y_{16}$, and $\tilde Y_{20}$, and taking
$\tilde Y_{12}$ to be fixed at the computed value.  In terms of the Jacobian
matrix $\bm J$ defined by
\begin{equation}
\bm J \equiv \pardiv{\bm F}{\bm Y}
\qquad
{\bm F} =
\left(
\begin{array}{c}
 F_0 \\ F_1 \\ F_2
\end{array}
\right)
\qquad
{\bm Y} =
\left(
\begin{array}{c}
 Y_\alpha \\ Y_{16} \\ Y_{20}
\end{array}
\right)
\label{4alpha1.5}
\end{equation}
a Newton--Raphson iteration step then corresponds to choosing an initial vector
$\bm Y$, computing the corresponding values of $\bm F$ and $\bm J$, solving the
matrix equation
\begin{equation}
 \bm J\, \delta \hspace{-0.5pt} {\bm Y} = -\bm F
\label{newty1.1}
\end{equation}
for the increment $\delta {\hspace{-0.5pt} \bm Y}$, and adding this increment to
the original $\bm Y$ to get a corrected $\bm Y \rightarrow \bm Y+\delta
\hspace{-0.5pt} {\bm Y}$.  The  corrected $\bm Y$ can then be used as the
starting point for a second iteration, and so on
until a convergence criterion is satisfied.

Because it involves a matrix solution, this method has the potential to spoil
the linear scaling with network size that is attractive about explicit methods
unless the matrix equations can be solved by means other than brute force. 
Because the matrices for large networks are expected to be sparse and
well-conditioned, it is likely that solution methods giving scaling not too
different from linear are possible, but this needs to be demonstrated for this
method. For our tests we shall solve these matrix equations using standard
matrix packages.

\subsubsection{Use Iteration to Reimpose Equilibrium Abundances}
\protect\label{abundanceIteration}

As an alternative to restoring abundance ratios, we may seek to reimpose
equilibrium by requiring that the individual abundances of all isotopes
participating in equilibrium be set to their equilibrium values
\eqnoeq{2body1.6} at the end of their numerical timestep. Suppose that at some
integration timestep there are $n$ reaction groups in equilibrium. There will be
some number of isotopes $N$ participating in these reaction groups, with some
isotopes participating in more than one equilibrated reaction group. Define a
vector $\bm Y$ of the $N$ distinct isotopes that are participating in at least
one equilibrated reaction group.  If we assume that both $\alpha +
\isotope{12}{C} \rightleftharpoons \isotope{16}{O}$ and  $\alpha +
\isotope{16}{O} \rightleftharpoons \isotope{20}{Ne}$ are in equilibrium for the
network \eqnoeq{4alpha1.0}, $N=4$ and $n=2$, and the components of $\bm Y$ are
\begin{equation}
 Y_i = \{ Y_\alpha, Y_{12}, Y_{16}, Y_{20}\}
\label{Yvec1.1}
\end{equation}
where  $Y_{12} = Y(\isotope{12}{C})$ and so on. Now define a vector $\bm F$ of
normalized differences between the value of $Y_i$ at the end of the numerical
timestep and its equilibrium value calculated from \eqnoeq{2body1.6} at the end
of the numerical timestep, and an entry imposing conservation of particle
number, and require that it vanish. For the above example, we obtain the vector
equation $\bm F = 0$, with the components of $\bm F$ given by
\begin{equation}
 F_i = \left\{
\tfrac{Y_\alpha - \bar Y_\alpha^{(1)}}{\bar Y_\alpha^{(1)}},
\tfrac{Y_{12} - \bar Y_{12}^{(1)}}{\bar Y_{12}^{(1)}},
\tfrac{Y_{16} - \bar Y_{16}^{(1)}}{\bar Y_{16}^{(1)}},
\tfrac{Y_\alpha - \bar Y_\alpha^{(3)}}{\bar Y_\alpha^{(3)}},
\tfrac{Y_{16} - \bar Y_{16}^{(3)}}{\bar Y_{16}^{(3)}},
\tfrac{Y_{20} - \bar Y_{20}^{(3)}}{\bar Y_{20}^{(3)}},
\Sigma_X -1
\right\} ,
\label{Yvec1.2}
\end{equation}
where $Y_i$ denotes the actual abundance of species $i$ at the end of the
numerical timestep and the computed equilibrium value of species $i$ at the end
of the timestep is $\bar Y_i^{(j)}$, with the index $i$ labeling the position in
the vector \eqnoeq{Yvec1.1} and the index $j$ labeling the reaction group for
which \eq{2body1.6} is solved. The first three entries in the vector
\eqnoeq{Yvec1.2} are associated with restoring equilibrium in reaction group 1
and the next three entries are associated with restoring equilibrium in reaction
group 3.  Notice that the abundances $Y_\alpha$ and $Y_{16}$ appear more than
once (twice each, for this example) in the vector $\bm F$.  This is because
these isotopes appear in  reaction group 1 and in reaction group 3, both of
which are assumed to be in equilibrium. For the last entry $\sum_X-1$ is defined
by \eq{4alpha1.4}, and is a requirement that total nucleon number be conserved
in the equilibrium restoration step.

Thus restoration of equilibrium at the end of the timestep, subject to
conservation of particle number, corresponds to solving $\bm F(\bm Y) = 0$ for
the vector $\bm Y$ (note that this vector contains only those isotopes that are
part of reaction groups in equilibrium, not all of the isotopes in the network).
As outlined in the previous section, the equation $\bm F(\bm Y)=0$ can be solved
iteratively for $\bm Y$ by choosing an initial guess for $\bm Y$, computing $\bm
F(\bm Y)$, solving the matrix equation \eqnoeq{newty1.1} for the increment
$\delta \bm Y$, computing the improved $\bm Y \rightarrow \bm Y + \delta\bm Y$,
and then repeating until a convergence tolerance is satisfied. Since the
equilibrium abundances for the isotopes are likely to change very slowly in
successive timesteps, this Jacobian matrix may be expected to be almost constant
between two successive timesteps, and it will not change in successive
Newton--Raphson iterations since the equilibrium abundances are constant within
a given timestep.

As for the method described in the previous section, a Newton--Raphson
iteration on matrix equations is employed, so it will be highly desirable to
solve these equations by means that retain the nearly linear scaling of the
explicit method. From the discussion we see that for large networks with many
reaction groups in equilibrium the matrices will be extremely sparse and only
slightly changed in successive timesteps (and unchanged in successive
Newton--Raphson iteration steps within a given timestep).  We may expect that
this structure lends itself to  solutions with favorable scaling behavior in
large networks, but that has not been investigated yet.

\subsubsection{Reimpose Equilibrium Abundances by Averaging}
\protect\label{abundanceAverage}

The methods described in the previous two sections for restoring equilibrium at
the end of numerical timesteps suffer from a certain level of redundancy because
we have seen that in partial equilibrium the isotopic abundances in a reaction
group are not independent but evolve according to a single timescale given by
\eq{2body1.8} (see the general discussion in \S\ref{progvar}). Thus, within a
single reaction group specification of the equilibrium abundance of any one
isotope, or of the progress variable associated with the reaction group,
specifies the equilibrium abundance of all species in the group.  Furthermore,
within a single reaction group the evolution of the species in the group to
equilibrium naturally conserves particle number, by virtue of constraints such
as those of \eq{2body1.1} that are listed for all five reaction group classes in
Appendix \ref{RGclassificationApp}.

Let us exploit this by working with the progress variable $\lambda_i$ from each
reaction group. If all the reaction groups were independent, then we could
restore equilibrium for the progress variables for $n$ reaction groups in
equilibrium simply by requiring
\begin{equation}
\bm F = \left(
\begin{array}{c}
 \lambda_1 - \bar\lambda_1
\\
\lambda_2 - \bar\lambda_2
\\
\vdots
\\
\lambda_n - \bar\lambda_n
\end{array}
\right)
=0
\label{progF1.1}
\end{equation}
where $\bar\lambda_i$ denotes the equilibrium value of $\lambda_i$ computed from
\eq{2body1.6} and relations like \eq{2body1.7}. Once the equilibrium value of
$\lambda_i$ (or the abundance of any one of the isotopes in the reaction group)
is computed, the equilibrium values for all other isotopes in the group than
follow from constraints like \eq{2body1.7} that are tabulated for all reaction
group classes in Appendix \ref{RGclassificationApp}.  No probability
conservation constraint is included in \eq{progF1.1} because each reaction
group considered in isolation conserves particle number automatically in the
evolution to equilibrium.  The solution of \eq{progF1.1} is then trivial,
consisting of setting $\lambda_i = \bar\lambda_i$ for all $i$.

The simple considerations of the preceding paragraph cannot be used in the form
presented: the reaction groups are generally {\em not} independent, because
isotopes of the network appear typically in more than one reaction group, as we
have discussed in \S\ref{complexityRealNet}. The equilibrium condition implies
that the equilibrium abundances of a specific isotope computed for each of the
reaction groups in which it is a member must be similar, but exact equality
generally will not hold because of the finite tolerance $\epsilon_i$ used to
test for equilibrium in \eq{2body1.8b}.   Thus, we restore equilibrium for each
isotope participating in partial equilibrium at the end of a timestep by
replacing its computed abundance with its equilibrium value averaged over all
equilibrated reaction groups in which it participates. If the species $i$ is a
member of $M$ equilibrated reaction groups, its restored equilibrium value at
the end of the timestep is 
\begin{equation} 
\bar Y_i = \frac 1M \sum_j^M \bar
Y_i^{(j)}. \label{progF1.4} 
\end{equation} 
where the equilibrium abundances $\bar Y_i^{(j)}$ are
computed for each of the reaction groups from \eq{2body1.6}.

There is one further issue:  although the evolution to equilibrium for
individual reaction groups conserves particle number, the averaging procedure of
\eq{progF1.4} over multiple reaction groups will introduce a fluctuation in the
particle number since the chosen value of $\bar Y_i$ computed from the average
will generally differ from the individual $\bar Y_i^{(j)}$ that were computed
conserving particle number.  This difference will be very small in any one
timestep, but can accumulate to unacceptable failure of particle number
conservation for a long integration. Thus, for each timestep, after equilibrium
has been restored through \eq{progF1.4} we recompute the total particle number
in the network (now summing over all isotopes, whether participating in
equilibrium or not), compare it with the total particle number at the beginning
of the timestep, and rescale all $Y_i$ in the network by a multiplicative factor
that restores the particle number to its initial value.

We have shown in a variety of comparisons that this procedure for restoring
equilibrium  works very well, giving results (isotopic abundances and
timestepping profile) that are essentially identical to those of the iterative
matrix algorithms discussed in the previous two sections.  The advantage of the
present approach is simplicity and automatically-favorable scaling with network
size. The restoration of equilibrium involves only a summation over reaction
groups to compute averages, followed by a multiplicative renormalization by a
scalar. Thus there are no matrix inversions and no Newton--Raphson iterations,
so this algorithm for restoring equilibrium is very simple to implement and
should scale linearly with the size of the network, straight out of the box. All
of the following examples were computed using this averaging algorithm.

\section{\label{ss:asyTimestep} A Simple Adaptive Timestepper}

For testing the algorithms described here an adaptive timestepper has
been employed that is described in more detail in Refs.\
\cite{guidAsy,guidQSS,guidJCP}. This timestepper is far from optimized
but it leads to stable and accurate results for the varied
astrophysical networks that have been tested.  Thus it is adequate for our
primary task here, which is to establish whether explicit partial
equilibrium methods can even
compete with implicit methods for stiff networks.

\section{\label{sh:toyModel} Toy Models for Partial Equilibrium}

As a first illustration of using the partial equilibrium methods developed in
the preceding sections, we take a very simple thermonuclear network
including only three isotopes, \isotope{4}{He} ($\alpha$-particles), 
\isotope{12}{C}, and \isotope{16}{O}, connected by the reactions
$
3\alpha \rightleftharpoons \isotope{12}{C}
$ 
and
$
\alpha + \isotope{12}{C} \rightleftharpoons \isotope{16}{O}.
$
Thus we have a total of four reactions in two reaction groups, with $3\alpha
\rightleftharpoons \isotope{12}{C}$ belonging to reaction group class C and
$\alpha + \isotope{12}{C} \rightleftharpoons \isotope{16}{O}$ belonging to
reaction group class B. The partial equilibrium formulas for both cases have
been summarized in  \ref{RGclassificationApp}. This is an extremely
simple network but we shall see that it demonstrates in a rather transparent
way most of the essential features of a partial equilibrium calculation.

Figure \ref{fig:3alphaTimesteps}%
 \putfig
     {3alphaTimesteps}
      {0pt}
      {\figdn}
      {1.20}
     {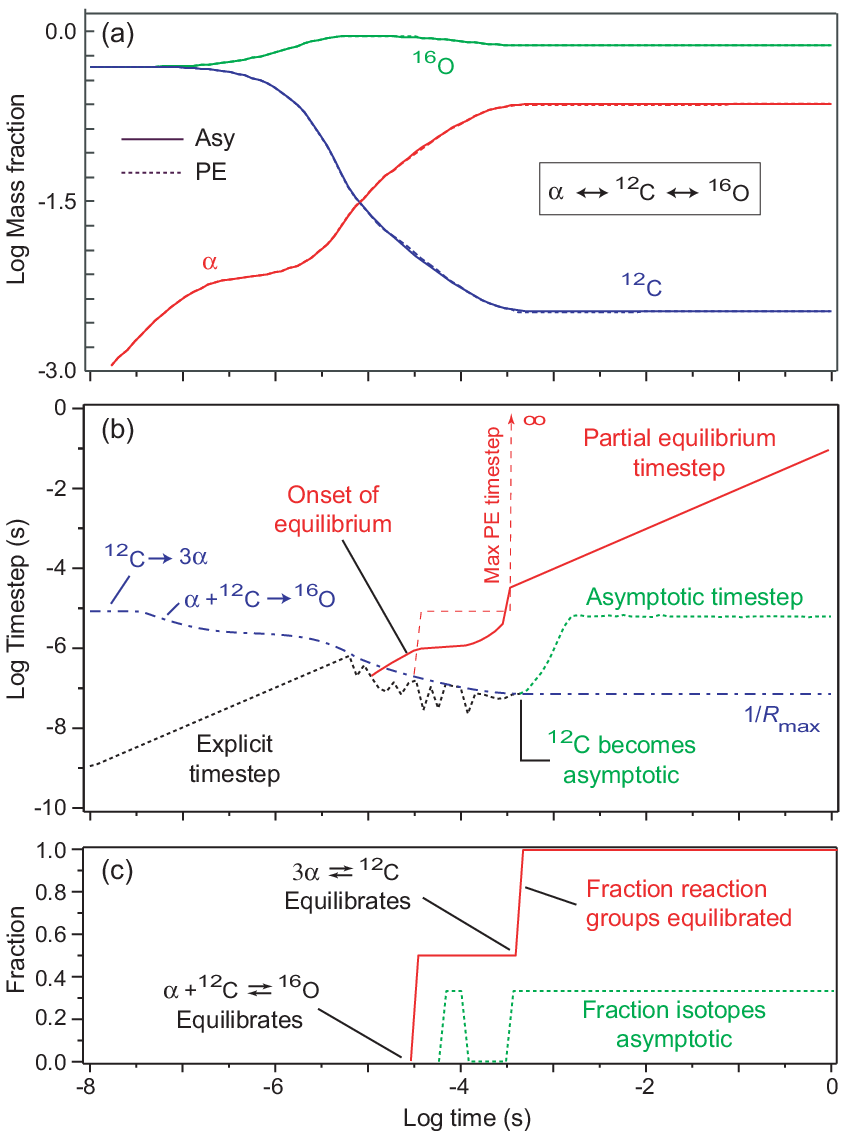}    
{The effect of imposing partial equilibrium on integration timesteps for the
network $\alpha \rightleftharpoons \isotope{12}{C} \rightleftharpoons
\isotope{16}{O}$ at a constant temperature of $5\times 10^9$ K and constant
density $1\times 10^8 \units{g\,cm}^{-3}$, assuming initial mass fractions of
$0.50$ for \isotope{12}{C} and \isotope{16}{O}.  Reaction rates were taken from
standard compilations \cite{raus2000}. (a) Mass fractions as a function of time.
The results are almost indistinguishable between the purely asymptotic
calculation (Asy) and the  asymptotic plus  partial equilibrium calculation
(PE). (b) The purely-explicit integration timestep (black dotted), the
asymptotic timestep (green dotted), the partial equilibrium timestep (solid
red), the approximate maximum explicit timestep for the asymptotic calculation
(dash-dotted blue), and the maximum stable timestep for the partial
equilibrium calculation (dashed red). (c) The fraction of isotopes that are
asymptotic and the fraction of reaction groups that are equilibrated as a
function of time. The dotted green curves in the middle and bottom figures refer
to the results one obtains with the explicit asymptotic method if partial
equilibrium is not imposed.}
illustrates a simulation with this network at a constant temperature of
$5\times 10^9$ K and constant density $1\times 10^8 \units{g\,cm}^{-3}$. This
calculation employs the explicit asymptotic method \cite{guidAsy}, but tries to
impose partial equilibrium  if the abundances of all reactions in a reaction
group are within $0.01$ of their equilibrium abundances.   The dash-dotted blue
curve labeled $1/R\tsub{max}$  estimates the maximum stable fully-explicit
timestep as the inverse of the current fastest rate in the network. This curve
is set by the $\isotope{12}C \rightarrow 3 \alpha$ reaction initially, but by
the $\alpha + \isotope{12}C \rightarrow \isotope{16}O$ reaction after $\log t
\sim -7.5$. 

First consider a purely asymptotic approximation. At the beginning the
maximum accurate timestep is smaller than the maximum stable explicit timestep,
so the algorithm takes explicit timesteps with a size set by accuracy and not
stability considerations (which we arbitrarily cap for this example at
$\diffelement t\sim 0.1 t$ to ensure accuracy). Near $\log t = -5.2$ the
explicit timestep becomes equal to the maximum stable explicit timestep
(intersection of dash-dotted blue and dotted black curves), which is decreasing
with time at this point because the rate for $\alpha +
\isotope{12}C \rightarrow \isotope{16}{O}$ is increasing with time. No
isotopes are yet asymptotic, so the explicit timestep begins to decrease in
order to remain below the dash-dotted blue curve and thus maintain stability,
with the fluctuations in the timestep curve representing the attempts by the
timestepping algorithm to increase the timestep being thwarted by the stiffness
instability. But around $\log t = -3.5$ one of the isotopes (\isotope{12}C)
becomes asymptotic and the timestep begins to increase rapidly over the explicit
stability limit. However, the explicit asymptotic algorithm is unable to take
large enough timesteps to make more isotopes asymptotic and the asymptotic
timestep saturates at late times near $\diffelement t \sim 1\times 10^{-5}$ s
(dotted green curve).

On the other hand, for the asymptotic plus PE calculation we proceed  as for the
purely asymptotic calculation until at  $\log t \sim -4.4$ the $\alpha +
\isotope{12}{C} \rightleftharpoons \isotope{16}{O}$ reaction is determined by
the network to satisfy the partial equilibrium condition and the net flux from
this pair of reactions is removed from the numerical integration by the PE
algorithm. Because of the removal of these fast components, the maximum stable
timestep for the PE calculation (dashed red curve in \fig{3alphaTimesteps}(b),
corresponding to the inverse of the fastest timescale remaining in the numerical
integration) begins to increase relative to that for the asymptotic calculation
(dash-dotted blue curve).  In response, the PE integration step size (solid red
curve in \fig{3alphaTimesteps}(b)) also begins to increase until at $\log t \sim
-3.4$ it is about 100 times larger than the purely asymptotic timestep. At this
point, the network determines that the $3\alpha \rightleftharpoons
\isotope{12}{C}$ reaction group also satisfies the partial equilibrium condition
and removes this pair of reactions from the numerical integration too. Now,
since there are only two equilibrium reaction groups in our simple network and
they have both been removed by the PE algorithm, the numerical integrator is
effectively solving the set of equations $d\bm Y = 0$.  Since all timescales
have now been removed from the numerical integration by the PE algorithm, there
is no stiffness instability at all and the maximum stable timestep (dashed red
curve) goes to infinity. 

Provided that the two reaction groups remain in equilibrium, we are now free to
take timesteps limited only by accuracy. Accordingly the PE timestep increases
rapidly and once again becomes equal to an upper limit of $\diffelement t \sim
0.1 t$  imposed by our arbitrary overall accuracy constraint (solid red curve).
Thus, after a network evolution time of one second the partial equilibrium
timestep is  approximately four orders of magnitude larger than that for the
explicit asymptotic algorithm, with the calculated isotopic abundances as a
function of time being indistinguishable in the two cases. This translates into
a total integration time that is 400 times faster for the asymptotic plus
partial equilibrium calculation versus the purely asymptotic calculation in the
example of \fig{3alphaTimesteps}.

A deeper understanding of the factors that determine the timesteps in
\fig{3alphaTimesteps} may be obtained by consulting
Fig.~\ref{fig:3alphaEquilTimescales},%
 \putfig
     {3alphaEquilTimescales}
      {0pt}
      {\figdn}
      {1.10}
     {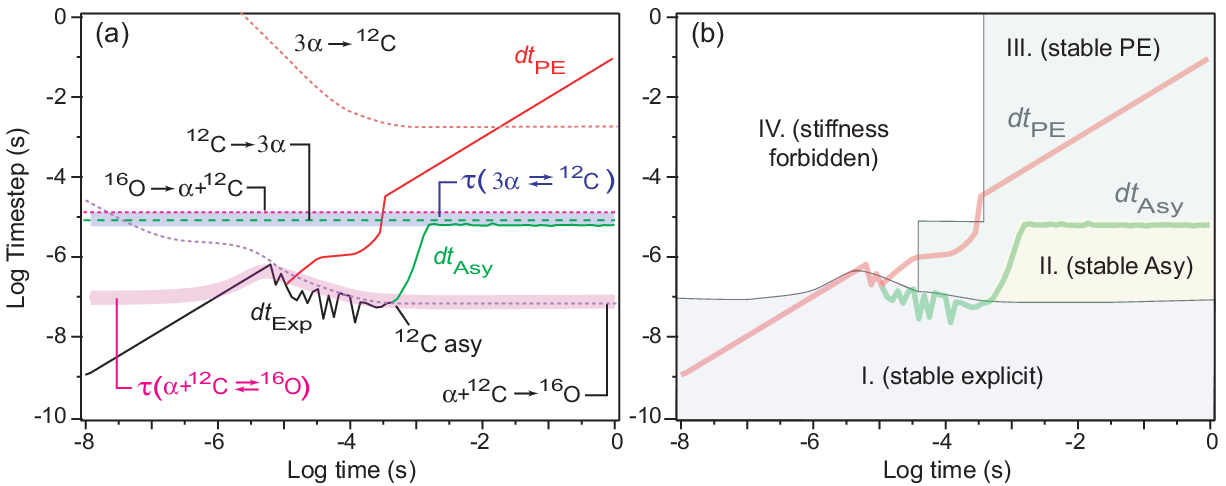}    
{(a)~Equilibration and reaction timescales for \fig{3alphaTimesteps}.
(b)~Stiffness stability domains implied by the equilibration timescales in part
(a).}
which displays all timescales relevant for the problem as a function of time.
Figure \ref{fig:3alphaEquilTimescales}(a) shows the timescales defined by the
inverse of the reaction rates for all reactions in the network (dotted and
dashed curves), the timescales $\tau$ of \eq{2body1.8} for establishing
equilibrium for the two reaction groups in this calculation (magenta and blue
bands), and the integration timestepping for various methods (solid curves). The
differential equations to be solved are those of Eqs.\
\eqnoeq{4alpha1.1a}--\eqnoeq{4alpha1.1d} with terms involving \isotope{20}{Ne}
removed:
\begin{eqnarray}
  \dot Y_{\alpha} \mwgalign= -\kforward 0 Y_\alpha^3
+ \kback 0 Y_{12} - \kforward 1 Y_\alpha Y_{12}  + \kback 1 Y_{16} 
\label{3alpha1.1a}
\\
\dot Y_{12} \mwgalign= \kforward 0 Y_\alpha^3 - \kback 0 Y_{12}
- \kforward 1 Y_\alpha Y_{12} +\kback 1 Y_{16} 
\label{3alpha1.1b}
\\
\dot Y_{16} \mwgalign= \kback1 Y_\alpha Y_{12} -\kback1 Y_{16}
-\kforward3 Y_\alpha Y_{16} 
\label{3alpha1.1c}
\end{eqnarray}
In the explicit asymptotic calculation (where we ignore the role of the
equilibrium timescales denoted by the magenta and blue bands) we see that the
timestep is rigidly limited by the inverse rate for
$
\alpha + \isotope{12}{C} \rightarrow \isotope{16}{O}
$
(dashed purple curve) until  \isotope{12}{C} becomes asymptotic (labeled
\isotope{12}{C} asy).  This partially lifts the timestepping restriction because
\eq{3alpha1.1a} and its source terms for
$
\alpha + \isotope{12}{C} \rightleftharpoons \isotope{16}{O}
$ 
are then removed from the numerical integration.  However, since no other
isotopes become asymptotic, the numerical integration is unable to get past the
ceiling imposed by the timescales associated with
 $
 \isotope{12}{C} \rightarrow 3\alpha
$
and  
$
 \isotope{16}{O} \rightarrow \alpha + \isotope{12}{C} 
$
that remain in Eqs.~\eqnoeq{3alpha1.1b}--\eqnoeq{3alpha1.1c} and the asymptotic
timestep saturates around $10^{-5}$ s at late times.

In contrast,  the partial equilibrium method is able to remove all timescales in
all three equations associated with the reaction group $\alpha + \isotope{12}{C}
\rightleftharpoons \isotope{16}{O}$ when the timestep $dt$ is comparable to the
magenta band denoting
$
\tau(\alpha + \isotope{12}{C} \rightleftharpoons \isotope{16}{O})
$.  
This complete removal of a set of fast timescales replaces the original problem
with one that is less stiff.  That permits the PE method to increase its
timestep sufficiently that $dt$ quickly reaches the equilibrium timescale
associated with 
 $
3\alpha \rightleftharpoons \isotope{12}{C}
$
(blue band) and removes completely all timescales associated with this
reaction group too, leaving an integration problem having
no stiffness limitation on the explicit timestep.  

An even simpler picture of the relationship between asymptotic and partial
equilibrium approximations emerges if we consider the evolution of a single
reaction group. In Fig.\ \ref{fig:RGequil}(a),%
 \putfig
     {RGequil}
      {0pt}
      {\figdn}
      {0.73}
     {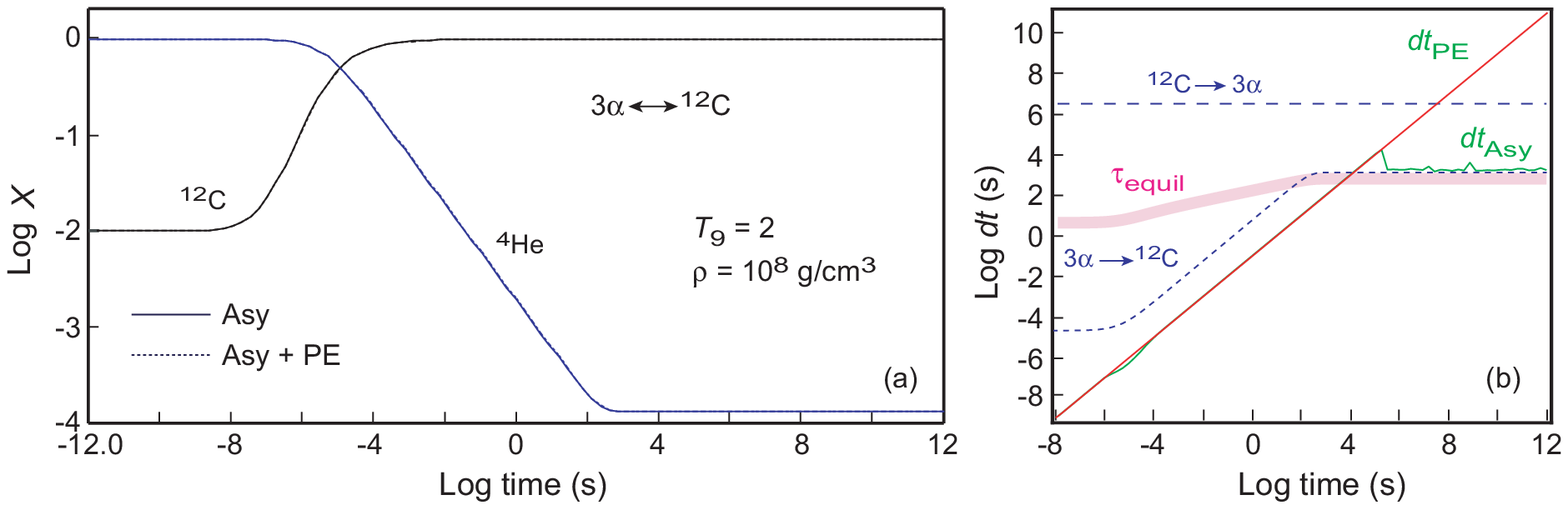}    
{Equilibration of the single reaction group $3\alpha \rightleftharpoons
\isotope{12}{C}$. (a)~Mass fractions in asymptotic and asymptotic plus partial
equilibrium approximation. (b)~Timescales and integration timesteps.  The dashed
curves are the timescales set by the inverses of the reaction rates; the dotted
red curve is the equilibration timescale \eqnoeq{2body1.8} for the reaction
group. The solid green curve is the integration timestepping for an asymptotic
approximation and the solid red curve is the timestepping for an asymptotic plus
partial equilibrium approximation.}
by evolving a single reaction group consisting of the triple-alpha reaction and
its inverse, we see clearly the role of the individual reaction timescales and
the timescale for approach to equilibrium of $3\alpha \rightleftharpoons
\isotope{12}{C}$ in setting the timestepping for asymptotic integration. For
accuracy in this example, we have arbitrarily limited the integration timestep
to $dt \le 0.1 t$.  As illustrated in \fig{RGequil}(b), initially, the largest
stable fully-explicit timestep is governed by the inverse of the rate for the
$3\alpha \rightarrow \isotope{12}{C}$ reaction.  Until $\log t \sim 4$, the
largest accurate timestep lies below this timescale, so there is no stiffness
limitation on the explicit integration. At around $\log t \sim 4$ the
\isotope{4}{He} ($\alpha$ particle) becomes asymptotic and the asymptotic
timestep begins to exceed the limit set by the $3\alpha \rightarrow
\isotope{12}{C}$ reaction by a small amount.  However, the other isotope in the
network, \isotope{12}{C}, never satisfies the asymptotic condition over the
entire range of integration. Thus, the asymptotic method is never able to remove
completely the stiffness timescales set by the $3\alpha \rightleftharpoons
\isotope{12}{C}$ reactions (since they will remain at all times in the
differential equation for \isotope{12}{C}, which must be integrated
numerically). Thus the asymptotic-method integration timestep saturates at late
times near the timescale set by the inverse of the rate for $3\alpha
\rightarrow \isotope{12}{C}$.

On the other hand, in the asymptotic plus PE calculation the reaction group
$3\alpha \rightleftharpoons \isotope{12}{C}$ becomes equilibrated according to
the criteria of \eq{2body1.8b} at about the time the equilibrium timescale for
the reaction group (shown as the magenta band) crosses the $\diffelement t$
curve near $\log t \sim 3$. Since the reaction group is now assumed to be in
equilibrium, all flux terms associated with both $3\alpha \rightarrow
\isotope{12}{C}$ and $\isotope{12}{C} \rightarrow 3\alpha$ are removed from the
numerical integration and the numerical integrator must solve the system
$\diffelement Y_{\alpha}/\diffelement t = \diffelement Y_{12}/\diffelement t =
0$. Since all timescales have been removed from the network, the corresponding
integration timestep is set only by accuracy criteria.  As a result, the partial
equilibrium integration in \fig{RGequil} is more than 300,000 times faster than
the asymptotic integration of the same system, with essentially identical
results.

Similar considerations apply in more complex networks and it is useful to view
the timescales of \fig{3alphaEquilTimescales}(a) or \fig{RGequil}(b) as
establishing stiffness stability domains in the $dt$ versus $t$ plane that are
illustrated in \fig{3alphaEquilTimescales}(b). Thus, a fully-explicit
integration is restricted by stability considerations to the lower domain marked
I, an asymptotic (or QSS) calculation is restricted to domains I and II, an
asymptotic or QSS plus partial equilibrium calculation is restricted to domains
I, II, and III, and all of these methods would be unstable in domain IV. 
However, as illustrated in \fig{3alphaEquilTimescales}(b), if we are
sufficiently clever we can push domain IV far enough into the upper left corner
of the $dt-t$ plane that it plays little practical role because the instability
domain corresponds to timesteps that would be undesirable from an accuracy point
of view.

The examples displayed in Figs.\ \ref{fig:3alphaEquilTimescales} and
\ref{fig:RGequil} are not very complex and it should be not at all surprising
that the description of the system becomes simple and the stable timesteps large
at later times, since the mass fraction curves become almost constant in this
region. But that is the whole point of the partial equilibrium approach:  near
equilibrium the system is complicated when viewed in terms of competing
independent reactions, but becomes simple when viewed in terms of groups of
reactions coming into equilibrium and thus bringing the whole system into
equilibrium. These examples illustrate for some very simple networks, but the
same principles will apply to the more complex networks to which we now turn.

\section{\label{sh:testsAlpha} Tests on Some Thermonuclear Alpha Networks}

We have tested the  partial equilibrium algorithm described in earlier sections
in a variety of thermonuclear alpha networks. In this section we give some
representative examples of those calculations. Because they are extremely
challenging reaction network problems to solve, we shall concentrate on examples
representative of conditions expected in Type Ia supernova simulations
(temperatures in the range $10^7$--$10^{10}$ K, densities in the range
$10^7$--$10^9$ g\,cm$^{-3}$, and initial equal mass fractions of \isotope{12}C
and \isotope{16}O). Such conditions lead quickly to high degrees of
equilibration in the approach to QSE (quasi-statistical equilibrium) and NSE
(nuclear statistical equilibrium), and are a stringent test of partial
equilibration methods. Our long-term goal is application of these methods to
larger networks, but an alpha network provides a highly-stiff test system  that
is small enough to allow significant insight into how the algorithm functions.
We also note as a practical matter that the most ambitious published
calculations for thermonuclear networks coupled to hydrodynamical simulations
have employed alpha networks. 

The reactions and corresponding reaction groups used in the calculations are
displayed in Table \ref{tb:alphaNetworkFullReverse}.
\begin{table}
\caption{\label{tb:alphaNetworkFullReverse}Reactions of the alpha network for
partial equilibrium calculations. The reverse reactions such as
$\isotope{20}{Ne} \rightarrow \alpha+\isotope{16}{O}$  are photodisintegration
reactions, $\gamma + \isotope{20}{Ne} \rightarrow \alpha+\isotope{16}{O}$, with
the photon $\gamma$ suppressed in the notation.}
\begin{indented}
\item[]\begin{tabular}{@{}cclc}
\br
Group & Class & Reactions & Members\\
\mr
            1 &
            C &
            $3\alpha\rightleftharpoons\isotope{12}{C}$ &
            4
	\\
	2 &
            B &
            $\alpha+\isotope{12}{C}\rightleftharpoons\isotope{16}{O}$ &
            4
\\
	3 &
            D &
           
$\isotope{12}{C}+\isotope{12}{C}\rightleftharpoons\alpha+\isotope{20}{Ne}$ &
            2
\\
	4 &
            B &
            $\alpha+\isotope{16}{O}\rightleftharpoons\isotope{20}{Ne}$ &
            4
\\
	5 &
            D &
           
$\isotope{12}{C}+\isotope{16}{O}\rightleftharpoons\alpha+\isotope{24}{Mg}$ &
            2
\\
	6 &
            D &
           
$\isotope{16}{O}+\isotope{16}{O}\rightleftharpoons\alpha+\isotope{28}{Si}$ &
            2
\\
	7 &
            B &
            $\alpha+\isotope{20}{Ne}\rightleftharpoons\isotope{24}{Mg}$ &
            4
\\
	8 &
            D &
           
$\isotope{12}{C}+\isotope{20}{Ne}\rightleftharpoons\alpha+\isotope{28}{Si}$ &
            2
\\
	9 &
            B &
            $\alpha+\isotope{24}{Mg}\rightleftharpoons\isotope{28}{Si}$ &
            4
\\
	10 &
            B &
            $\alpha+\isotope{28}{Si}\rightleftharpoons\isotope{32}{S}$ &
            2
\\
	11 &
            B &
            $\alpha+\isotope{32}{S}\rightleftharpoons\isotope{36}{Ar}$ &
            2
\\
	12 &
            B &
            $\alpha+\isotope{36}{Ar}\rightleftharpoons\isotope{40}{Ca}$ &
            2
\\
	13 &
            B &
            $\alpha+\isotope{40}{Ca}\rightleftharpoons\isotope{44}{Ti}$ &
            2
\\
	14 &
            B &
            $\alpha+\isotope{44}{Ti}\rightleftharpoons\isotope{48}{Cr}$ &
            2
\\
	15 &
            B &
            $\alpha+\isotope{48}{Cr}\rightleftharpoons\isotope{52}{Fe}$ &
            2
\\
	16 &
            B &
            $\alpha+\isotope{52}{Fe}\rightleftharpoons\isotope{56}{Ni}$ &
            2
\\
	17 &
            B &
            $\alpha+\isotope{56}{Ni}\rightleftharpoons\isotope{60}{Zn}$ &
            2
\\
	18 &
            B &
            $\alpha+\isotope{60}{Zn}\rightleftharpoons\isotope{64}{Ge}$ &
            2
\\
	19 &
            B &
            $\alpha+\isotope{64}{Ge}\rightleftharpoons\isotope{68}{Se}$ &
            2
        \\
\br       
\end{tabular}
\end{indented}
\end{table}
We use the standard REACLIB library \cite{raus2000} for all rates except for
three of the heavy-ion reactions (corresponding to reaction groups 3, 5, and 6
in the table) which have inverse reactions that are absent from the REACLIB
compilation and were taken from Ref.\ \cite{JINA}. For typical Type Ia supernova
conditions the rates for the capture and photodisintegration reactions in Table
\ref{tb:alphaNetworkFullReverse} become comparable and as large as $\sim
10^{10}-10^{12}\units{s}^{-1}$.  The corresponding maximum stable fully-explicit
integration timestep will be of order the inverse of the maximum rate in the
network, so timesteps as short as $\sim 10^{-12}\units{s}$ may be required for
stability of a fully-explicit integration.  On the other hand, the
characteristic physical timescale for the primary Type Ia explosion mechanism is
of order one second.  Thus integration of the alpha network coupled to a
hydrodynamics simulation of a Type Ia explosion could require $10^{12}$ or more
explicit network integration timesteps, which is obviously not practical and 
indicates graphically that this is an extremely stiff system.

\subsection{Comparisons of Explicit and Implicit Integration Speeds}

We shall be comparing explicit and implicit methods using codes that are at very
different stages of development and optimization. Thus, they cannot simply be
compared directly with each other.  We assume that for codes at similar levels
of optimization the primary difference between explicit and implicit methods 
would be in the extra time spent in implicit-method matrix operations. Hence, if
the fraction of time spent on linear algebra is $f$ for an implicit code, we
assume that an explicit code at a similar level of optimization could compute a
timestep faster by a factor of $F = 1/(1-f)$. Factors $F$ have been
tabulated in Ref.\ \cite{guidAsy} for several networks. Then we may compare
roughly the speed of explicit versus implicit codes (possibly at different
levels of optimization) by multiplying $F$ by the ratio of implicit to explicit
integration steps required for a given problem.  This procedure has obvious
uncertainties, and likely underestimates the speed of an optimized explicit
versus optimized implicit code, but will give a useful lower limit on how fast
the explicit calculation can be relative to implicit methods.

\subsection{Constant Intermediate-Temperature, Low-Density Example}

A calculation for constant $T_9=5$ and density of $1\times 10^{7}
\units{g\,cm}^{-3}$ in an alpha network is shown in \fig{alpha507PE}.
 \putfig
     {alpha507PE}
     {0pt}
     {20pt}
     {0.75}
     {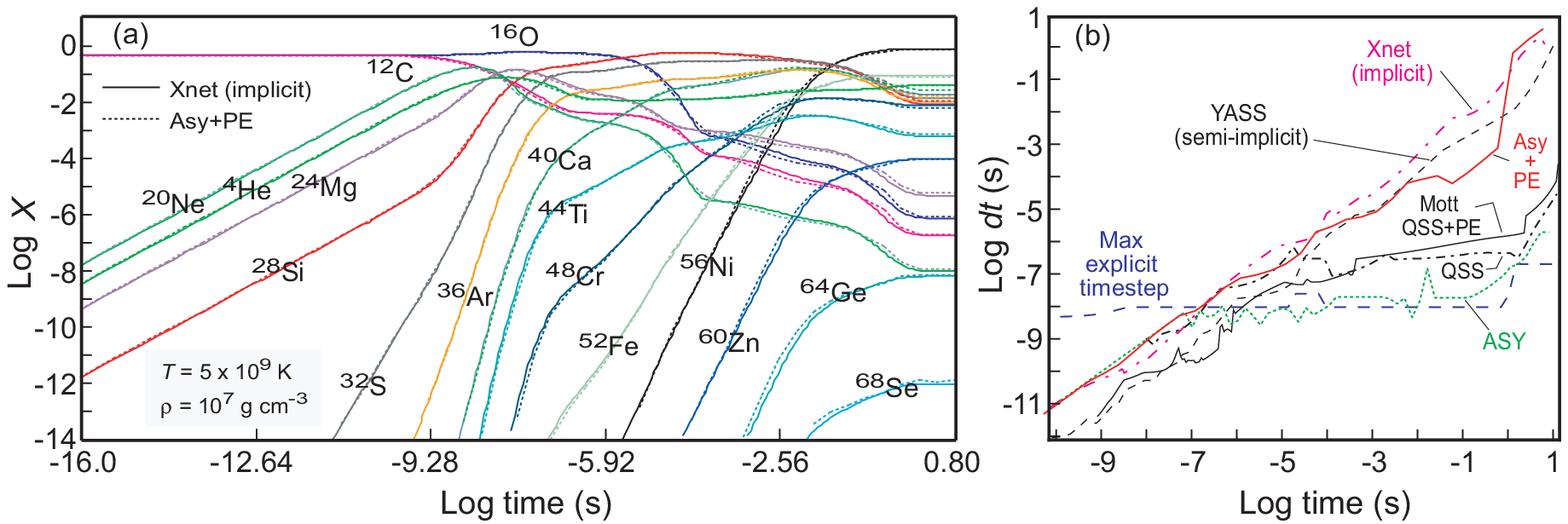}
{ Partial equilibrium calculation for constant $T_9=5$ and $\rho=1\times
10^{7}\units{g\,cm}^{-3}$ alpha network. (a)~Mass fractions. Solid lines are
implicit calculations (Xnet) and dashed lines are asymptotic + PE calculations.
(b)~Integration timesteps for several different integration methods. Magenta
dash-dot is the implicit code Xnet \cite{raphcode}, black dashed is the
semi-implicit code YASS \cite{YASS} (reproduced from from Ref.\ \cite{mott99}),
solid red is the current asymptotic plus partial equilibrium result, solid black
is the QSS plus partial equilibrium calculation reproduced from  Ref.\
\cite{mott99}, black dash-dot is a QSS calculation using the formalism of Ref.\
\cite{guidQSS}, dotted green is an asymptotic calculation using the formalism in
Ref.\ \cite{guidAsy}, and dashed blue is the estimated maximum stable
fully-explicit timestep. }
This network has 16 isotopes with 48 reactions connecting them, and a total of
19 reaction groups that can separately come into equilibrium. For this and all
other calculations
equilibrium conditions were imposed using \eq{2body1.8b},
with a constant value $\epsilon_i = 0.01$ for the tolerance parameter. The
calculated
asymptotic plus partial equilibrium (Asy + PE) mass fractions are compared with
those of a standard implicit code in \fig{alpha507PE}(a).  There are small
discrepancies in localized regions, especially for some of the weaker
populations, but for the most part the agreement is rather good.  Although we
shall generally display mass fractions down to $10^{-14}$ in our examples for
reference purposes, it is important to note that for reaction networks coupled
to hydrodynamics only larger mass fractions (those larger than say $\sim
10^{-2}-10^{-3}$) are likely to have significant influence on the
hydrodynamics. Thus, the largest discrepancies between mass fractions calculated
by PE and implicit methods in \fig{alpha507PE}(a) mostly imply uncertainties in
the total mass being evolved by the network of less than one part in a million,
which would be completely irrelevant in a coupled hydrodynamical simulation.

Timestepping for various integration methods is illustrated in
\fig{alpha507PE}(b), where there are several interesting features to discuss.  

\begin{enumerate}

\item
The PE timestepping is seen to compare rather favorably with standard
fully-implicit and semi-implicit codes. The PE calculation required 3941
integration steps while the implicit code Xnet required only 600 steps, but this
factor of 6.5 timestepping advantage is offset substantially by the expectation
that for a 16-isotope network an explicit calculation should be $\sim 3$ times
faster than the implicit calculation for each timestep \cite{guidAsy}.  Thus we
conclude that for fully-optimized codes the implicit calculation would be
perhaps several times faster for this example. The semi-implicit
timestepping curve was reproduced from another reference but a comparison of the
curves in
\fig{alpha507PE}(b) suggests that an optimized partial equilibrium code is
likely to be at least as fast as the semi-implicit YASS code for this example.
 \item 
The Asy + PE and implicit timestepping are seen to be many orders of magnitude
better at late times than the results of our purely asymptotic (Asy) and
purely quasi-steady-state (QSS) calculations, which are based on the
formalisms discussed in Refs.~\cite{guidAsy,guidQSS}. The reason is apparent
from \fig{mottCompare507fractionShort}. For times later than about $\log t =
-6$, significant numbers of reaction groups come into partial equilibrium,
which asymptotic and QSS methods are not designed to handle.
\item
The earlier application by Mott et al \cite{mott99} of a QSS plus
partial equilibrium calculation for this system  lags  orders
of magnitude behind the current implementation of asymptotic plus partial
equilibrium in timestepping at late times. In fact, the timestepping from Ref.\
\cite{mott99} is only a little better than that of the pure QSS results from the
present paper.

\end{enumerate}
 \putfig
     {mottCompare507fractionShort}
     {0pt}
     {\figdn}
     {0.40}
     {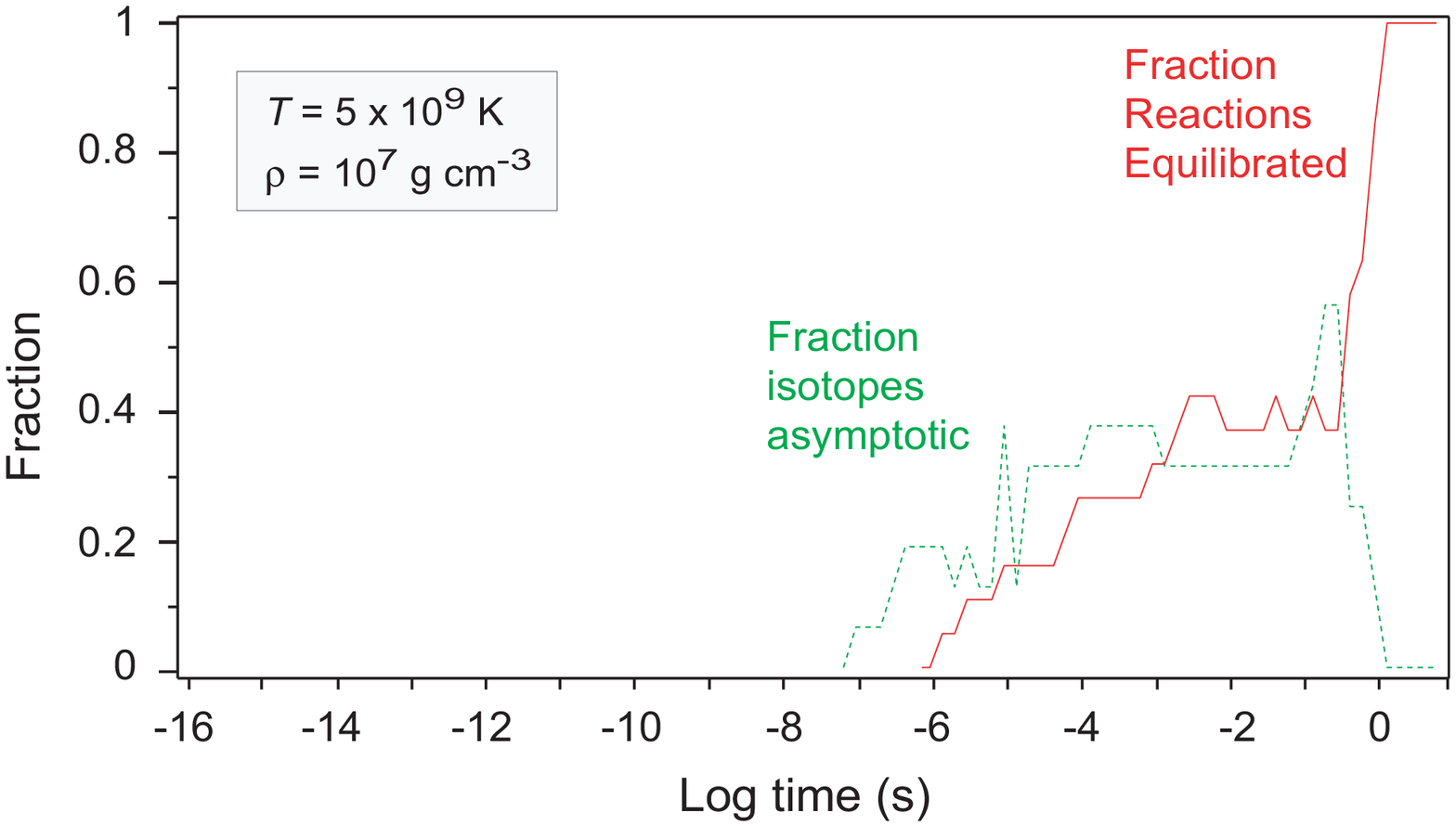}
{ Fraction of reactions that are treated as being in equilibrium  as a function
of integration time for the calculation of \fig{alpha507PE} (solid red curve). 
Also shown is the fraction of isotopes that become asymptotic in the PE
calculation (dotted green curve).}

\noindent
In \fig{tauPlotAlphaT9_5rho_1e7PEremove}%
 \putfig
     {tauPlotAlphaT9_5rho_1e7PEremove}
      {0pt}
     {20pt}
     {0.81}
     {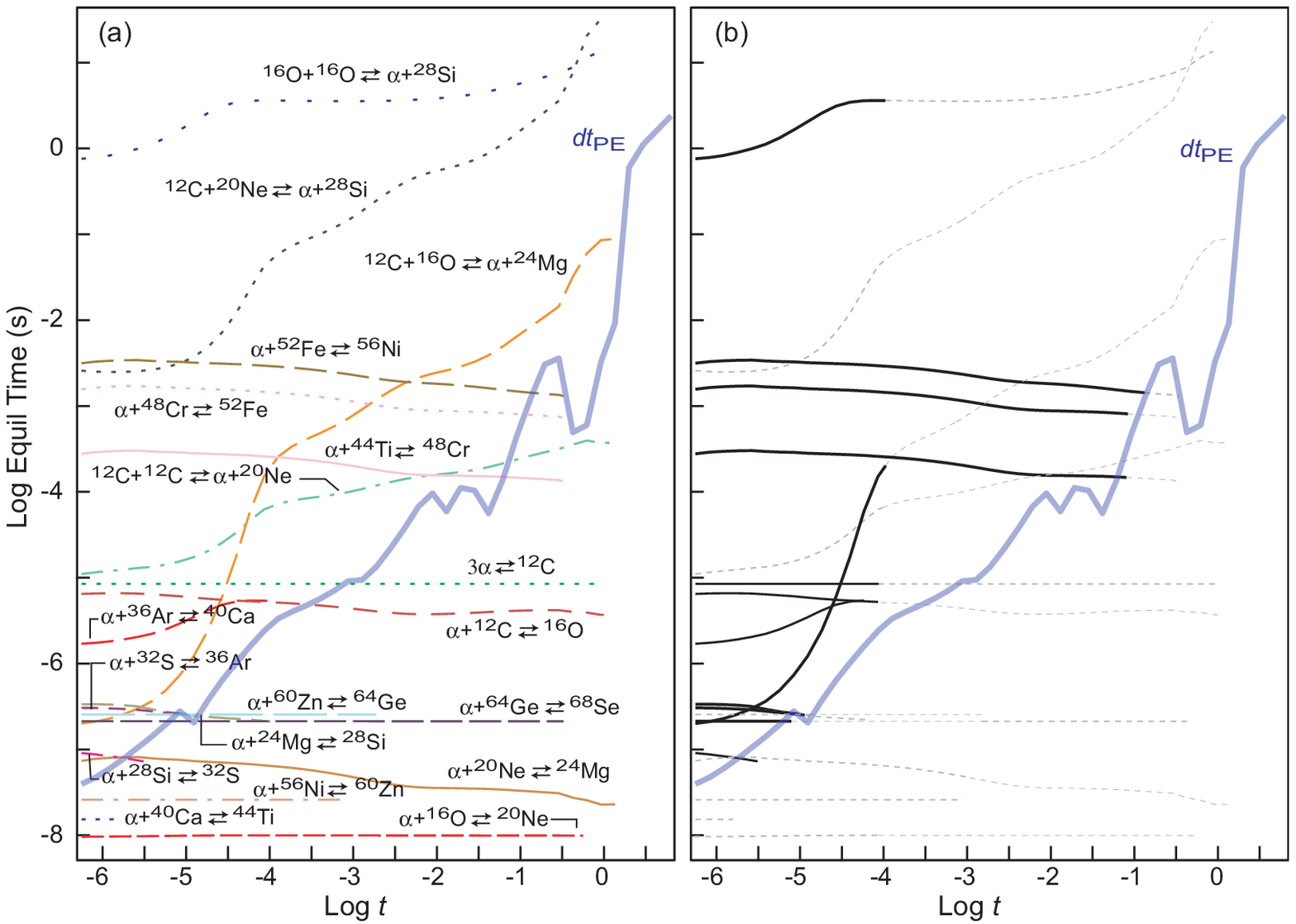}  
{ Timestepping and partial equilibrium timescales associated with the
calculation in \fig{alpha507PE}.  (a)~The equilibrium timescales, with the
reactions removed for times after they become equilibrated in the PE
calculation. (b)~As for part (a), but with labels suppressed and with time
ranges for which equilibrium reactions have been partially removed by the
asymptotic approximation indicated by dashed curves. The heavy, dark curves
indicate times for which the corresponding equilibrium timescale has been
removed neither by asymptotic nor PE approximations and therefore is fully
operative in producing stiffness. These curves clearly have a large influence
on the maximum timestep $dt\tsub{PE}$ that the PE calculation is able to take.}
we display the partial equilibrium timescales of \eq{2body1.8} that are
associated with the calculation of \fig{alpha507PE}, with the reactions
 that are removed by partial equilibrium or asymptotic considerations indicated
(compare with the simpler earlier example in \fig{3alphaEquilTimescales}). In
\fig{tauPlotAlphaT9_5rho_1e7PEremove}(a) each reaction has been removed from the
graph when it reaches equilibration.  In
\fig{tauPlotAlphaT9_5rho_1e7PEremove}(b) we also indicate in dashed gray those
reactions that have been at least partially removed from the numerical
integration by isotopes becoming asymptotic, or ranges of times for reactions in
which reactants have very small concentration and therefore make only small
contribution.  The complete removal of reactions by partial equilibration and
the partial removal of reactions by asymptotic conditions clearly play a
significant role in determining the envelope of the maximum stable and accurate
partial equilibrium timestep, as seen in
\fig{tauPlotAlphaT9_5rho_1e7PEremove}(b).

\subsection{Constant Higher-Temperature, Intermediate-Density Example}

A calculation for constant $T_9=7$ and density of $1\times 10^{8}
\units{g\,cm}^{-3}$ in an alpha network is shown in 
\fig{alphaT9_7rho1e8AsyXnetPE}.
 \putfig
     {alphaT9_7rho1e8AsyXnetPE}
     {0pt}
     {\figdn}
     {0.75}
     {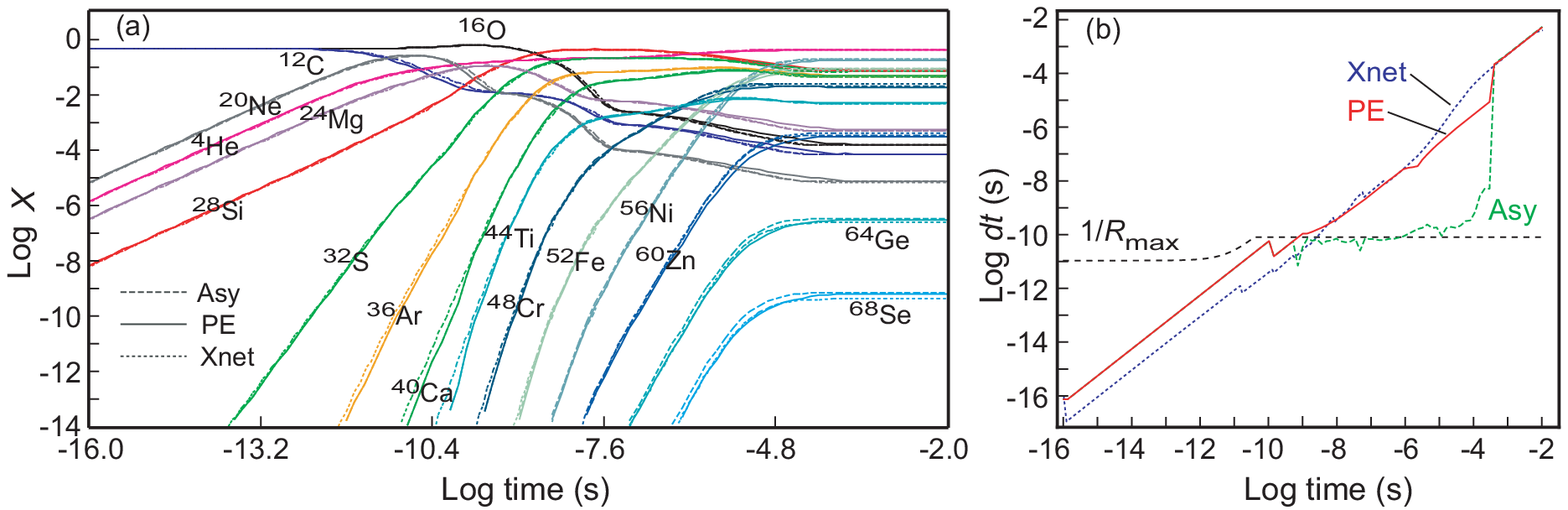}
{ Partial equilibrium calculation for constant $T_9=7$ and $\rho=1\times
10^{8}\units{g\,cm}^{-3}$ alpha network. (a)~Mass fractions. Solid curves are
asymptotic + PE, dashed curves are asymptotic, and dotted curves are implicit
(Xnet). (b)~Integration timesteps for the present asymptotic plus partial
equilibrium calculation (solid red), the implicit code Xnet \cite{raphcode}
(dotted blue), and the asymptotic method of Ref.\ \cite{guidAsy} (dashed green).
}
Again we see extremely good agreement among the mass fractions calculated by an
implicit method, the asymptotic method, and the asymptotic plus partial
equilibrium method.  The timesteps for the partial equilibrium and implicit
methods are quite similar, with the implicit method requiring 510 integration
steps and the partial equilibrium method 506 integrations steps. An optimized
explicit code can probably evaluate a timestep $\sim 3$ times faster than an
implicit code for a 16-isotope network \cite{guidAsy} and we conclude that for
similarly-optimized codes the partial equilibrium calculation would be slightly
faster.  In contrast, the asymptotic method lags far behind at late times,
because the equilibration becomes significant after $10^{-8}$ s, with the
fraction of reactions that are equilibrated reaching     100\% for times later
than $10^{-4}$ s. As a consequence, the integration of
\fig{alphaT9_7rho1e8AsyXnetPE} required more than two million purely-asymptotic
integration steps.

\subsection{Example with a Hydrodynamical Profile}

The preceding two examples employed alpha networks at extreme but constant
temperature and density. A partial equilibrium calculation using a
hydrodynamical profile with the dramatic temperature rise characterizing a
thermonuclear runaway in a Type Ia supernova simulation is illustrated in
\fig{torch47AlphaPE}.%
\putfig
{torch47AlphaPE}
{0pt}
{\figdn}
{0.72}
{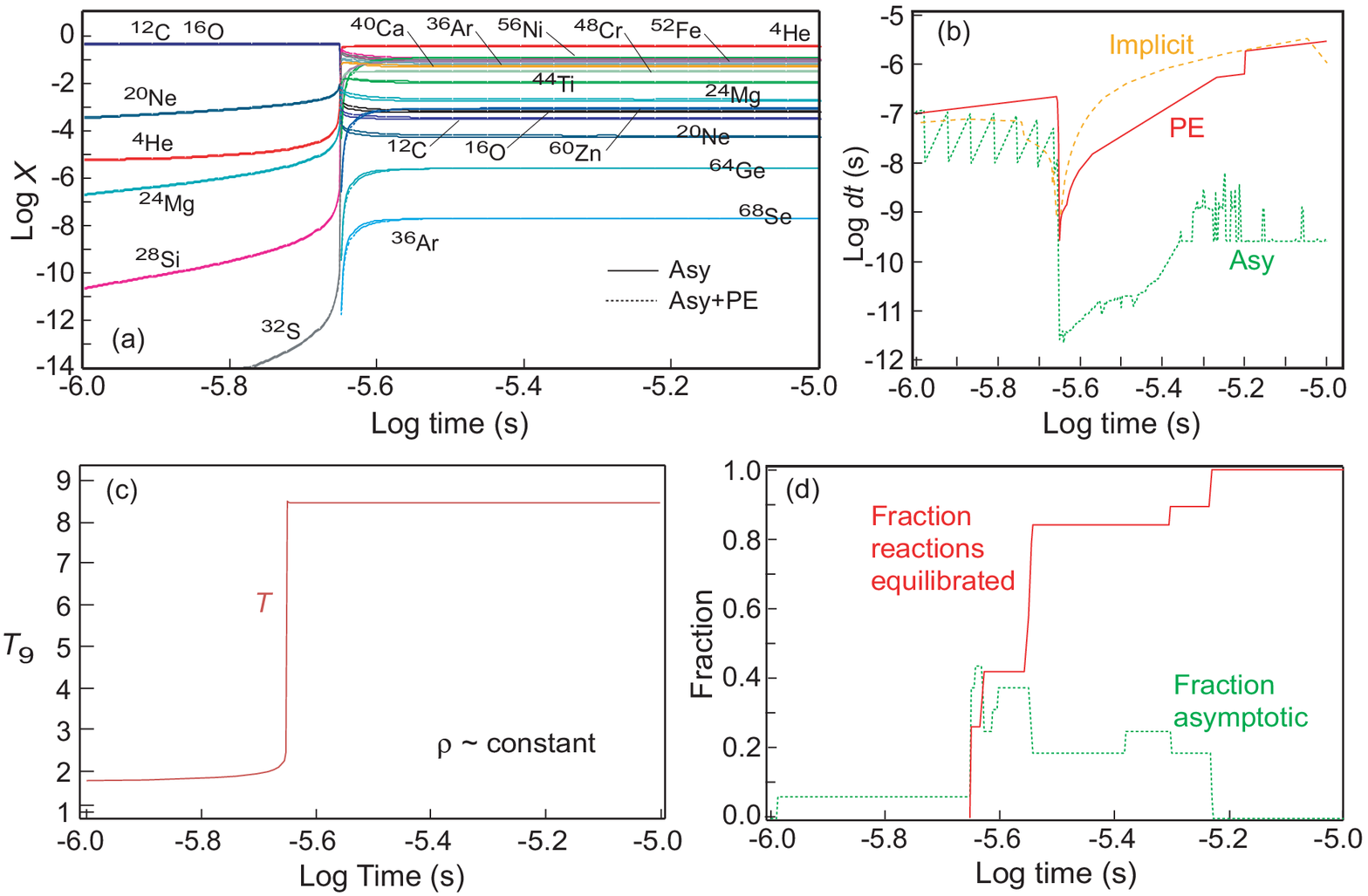}
{Asymptotic, PE, and implicit calculations with an alpha network for Type Ia
supernova conditions.  (a)~Mass fractions. (b)~Integration timesteps.
(c)~Hydrodynamic profile \cite{torch47}. (d)~Fraction of isotopes asymptotic and
partially-equilibrated reactions.}
The thermonuclear runaway in degenerate white-dwarf matter subjects the reaction
network to extreme conditions in this example: the difference between the
fastest and slowest reaction timescales is more than 11 orders of magnitude and
during the thermonuclear flash the temperature  increases by 6.6 billion K in
only $1.5 \times 10^{-7}$ s (a rate of temperature change corresponding to $4.4
\times 10^{16} \units {K/s}$). The asymptotic plus PE method required 712 steps
to integrate this problem, compared with 524 steps for the implicit calculation,
suggesting that an optimized explicit code would be about twice as fast as the
implicit code because it can compute a timestep about three times faster. In
contrast, the purely asymptotic calculation gave accurate results but required
almost 190,000 integration steps for the time range shown. 

As may be seen from \fig{torch47AlphaPE}(d), the partial equilibrium calculation
becomes completely equilibrated after $\log t \sim -5.2$ and after that time the
PE timesteps are limited only by accuracy. Choosing to restrict the subsequent
timesteps to $dt=0.3t$, continuation of the PE (or implicit) integration of
\fig{torch47AlphaPE} until the characteristic timescale for the supernova
explosion of $\sim 1$ s requires only a few tens of additional integration
steps. In contrast, if the asymptotic integrator continued to take the timesteps
it is taking at the end of the calculation in \fig{torch47AlphaPE}, it would
require more than a billion additional integration steps to reach the full
physical timescale for the supernova.

\subsection{Synopsis}

The three examples shown above are representative of a number of problems that
we have investigated with the new asymptotic plus partial equilibrium methods.
Our general conclusion is that for alpha networks under the extreme conditions
corresponding to a Type Ia supernova explosion the asymptotic plus partial
equilibrium algorithm is capable of giving timestepping that is orders of
magnitude better than asymptotic or QSS approximations alone when the system
approaches equilibrium, that these timesteps are often  competitive with current
implicit and semi-implicit codes, and that they are orders of magnitude faster
than previous attempts to apply such explicit methods to extremely stiff
thermonuclear networks.

\section{Improving the Speed of Explicit Codes}

The preceding discussion, and that of Refs.~\cite{guidAsy,guidQSS}, have assumed
that we can compare (approximately) the relative speed of explicit and
implicit methods that are presently implemented with different levels of
optimization by comparing the number of integration timesteps required to do a
particular problem with each method. The assumption of this comparison is that
once the matrix overhead associated with implicit methods is factored out, the
rest of the work required in a timestep (computing rates and fluxes, and so on)
is similar for explicit and implicit methods.  While this is useful for a first
estimation, it should not be viewed yet as a quantitative comparison of the
speed of implicit and explicit methods for two basic reasons.  (1)~The
current timestep routines for implicit methods and the explicit
methods presented here are at very different levels of algorithmic
optimization. (2)~Because the timestep routines are not yet equivalently
optimized, we have not made a systematic study of optimal timestepping
parameters for either the implicit or explicit methods.  Thus, the results
presented here should be interpreted as indicating that for the problems
investigated explicit methods can take stable and accurate timesteps that are
comparable to those of implicit methods.  Further work will be required to
determine whether implicit or explicit methods are faster and by how much for
specific problems.

There are at least three reasons why the integration speed for
explicit methods may potentially be even faster than that estimated by such a
procedure in this paper and in Refs.~\cite{guidAsy,guidQSS}.

First, in this paper we have emphasized partial equilibrium methods in
league with asymptotic approximations.  But in Ref.~\cite{guidQSS} we presented
evidence that quasi-steady-state (QSS) methods give results comparable to
asymptotic methods, but often with timesteps that can be as much as an order of
magnitude larger. Thus, we may expect that a QSS plus partial equilibrium
approximation could give even better timestepping than the examples shown here.
We have not yet investigated this possibility systematically.

Second, we have seen that the timestepping algorithm employed in this paper and
in Refs.~\cite{guidAsy,guidQSS} is not very optimized.  On the one hand, results
such as those exhibited in \fig{tauPlotAlphaT9_5rho_1e7PEremove}(b) suggest that
our timestepper is tracing qualitatively the explicit-integration stability
boundary.  But that same figure suggests that in many regions explicit timesteps
as much as an order of magnitude larger could still be stable. Thus, it is
possible that we have over-estimated the number of integration steps required by
our explicit methods in the examples discussed in this paper because of an
adequate but not optimized timestepper. This possibility also remains to be
investigated.

Third, and potentially of most importance, our demonstration that
algebraically-stabilized explicit methods may in fact be viable for large,
extremely-stiff networks implies new considerations with respect optimization.
For standard implicit methods, the most effective optimizations involve
improving the numerical linear algebra, since in large networks more than 90\%
of the computing time for implicit methods can be consumed by matrix inversions.
But with algebraically-stabilized explicit methods  the bulk of the computing
time for large networks is spent in computing the rates. Thus, the most
effective optimization for our explicit methods lies in increasing the speed
with which reaction rates are computed.  For example, can effective schemes to
compute rates that exploit modern multi-core architectures with GPU accelerators
be developed for realistic simulations of reaction networks coupled to
hydrodynamics? Such optimizations could be used for implicit integrations too,
but they will be less effective in increasing their speeds because in large
networks determining reaction rates is only a small part of the computing budget
of an implicit method. In contrast, we would expect our explicit integration
methods to gain speed substantially with faster methods to compute reaction
rates.

\section{\label{sh:extendFull} Extension to Larger Thermonuclear Networks}

As we have shown elsewhere, well away from equilibrium asymptotic and
quasi-steady-state approximations can compete with or even out-perform current
implicit codes in a variety of extremely stiff networks \cite{guidAsy,guidQSS}.
There are various important problems, such as the  nova and tidal supernova
examples discussed in Refs.\ \cite{guidAsy,guidQSS}, where the system does not
equilibrate strongly.  Even in problems where equilibrium is important overall,
there will be many hydrodynamical zones and timesteps for which equilibrium is
not very important. For these cases, the asymptotic and QSS approximations alone
may be viable. However, to compete across the board with implicit solvers, the
present paper makes clear that asymptotic and QSS approximations must be
supplemented by partial equilibrium methods to remain viable in those zones
where non-trivial equilibrium processes are present.

The examples of the previous section have shown that asymptotic plus partial
equilibrium methods can solve stiff thermonuclear alpha networks with
timestepping in the same ballpark as implicit or semi-implicit codes, and
accuracy sufficient for coupling  to fluid dynamics codes.  Although alpha
networks represent the current state of the art in coupling reaction networks to
hydrodynamics in such problems, our primary goal is to use these methods to
extend network sizes to physically more reasonable values (hundreds or thousands
of isotopes). Because the explicit methods that we are developing can compute a
timestep faster, and scale approximately linearly and therefore more favorable
with network size than implicit methods, their relative advantage grows with the
size of the network. Thus, for explicit methods to realize their full potential
we must extend the previous partial equilibrium examples to include very stiff
networks containing hundreds of reacting species. That work is underway and will
be reported in future publications, but in this section we make some general
remarks and report some preliminary results concerning these efforts.

\subsection{Relative Stiffness of Alpha and Realistic Thermonuclear Networks}

The primary difference (other than size) between an alpha network and a more
realistic thermonuclear network is that a more realistic network will typically
contain protons and neutrons.  The rates for reactions with protons and neutrons
are often many orders of magnitude higher than those induced by alpha particles;
as a result, realistic networks tend to contain much broader ranges of
characteristic timescales than alpha networks, and thus are often {\em much}
stiffer, as illustrated in \fig{maxRateAlphaVs150Hydro}.
\putfig
{maxRateAlphaVs150Hydro}
{0pt}
{\figdn}
{1.00}
{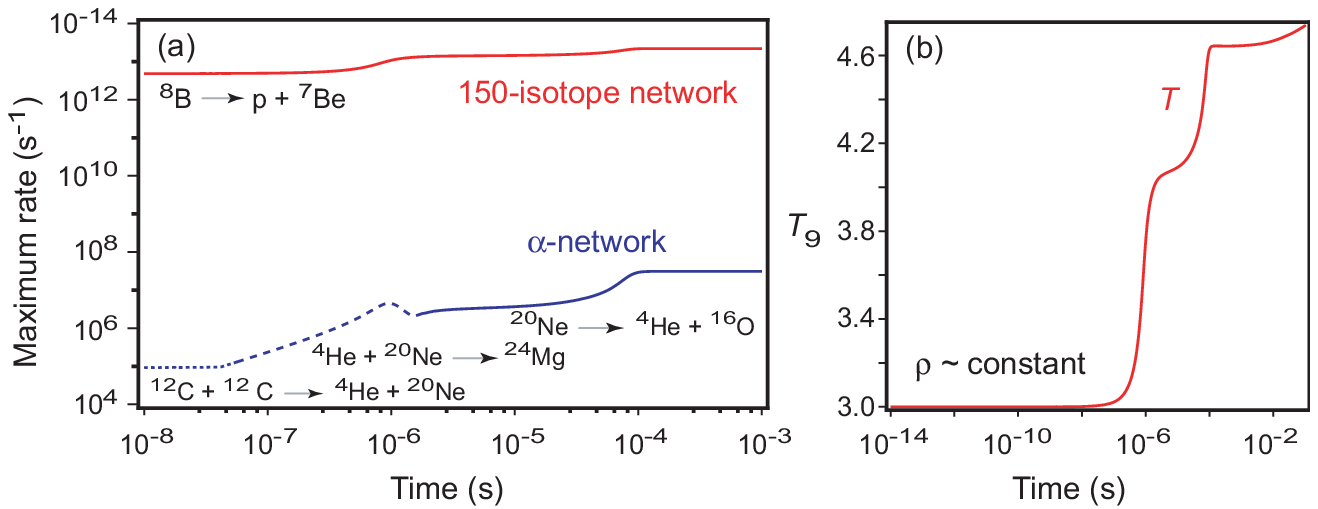}
{(a)~Maximum rates as a function of time in an alpha network and a 150-isotope
network for constant density and a hydrodynamic temperature profile given in
part (b). Rates were computed using the compilation in Ref.~\cite{raus2000}.
(b)~Hydrodynamical temperature profile corresponding to the calculation shown in
part (a); the density is approximately constant over this time range. This
profile is characteristic of evolution under Type Ia supernova conditions in a
single zone of a carbon--oxygen white dwarf with an initial $T_9 = 3$ and
$\rho=1\times 10^7 \units{g\,cm}^{-3}$.}
The fastest rates in the realistic network are seen to be 6--8 orders of
magnitude larger than those in the alpha network for this example. These very
fast reactions will tend to bring reaction groups into equilibrium rapidly.
Thus, we find that realistic networks are much stiffer than alpha networks, but
the most important effect for an explicit integration is not the increased
magnitude of the stiffness itself (which the asymptotic and QSS approximations
can handle in a manner competitive with implicit methods, as documented in
Refs.\ \cite{guidAsy,guidQSS}), but rather that these fast reactions quickly
manifest themselves in fast equilibration timescales, which asymptotic and QSS
algorithms cannot deal effectively with unless supplemented by a partial
equilibrium approximation. 

In \fig{frac150_alpha_508}%
 \putfig
     {frac150_alpha_508}
     {5pt}
     {\figdn}
     {0.78}
     {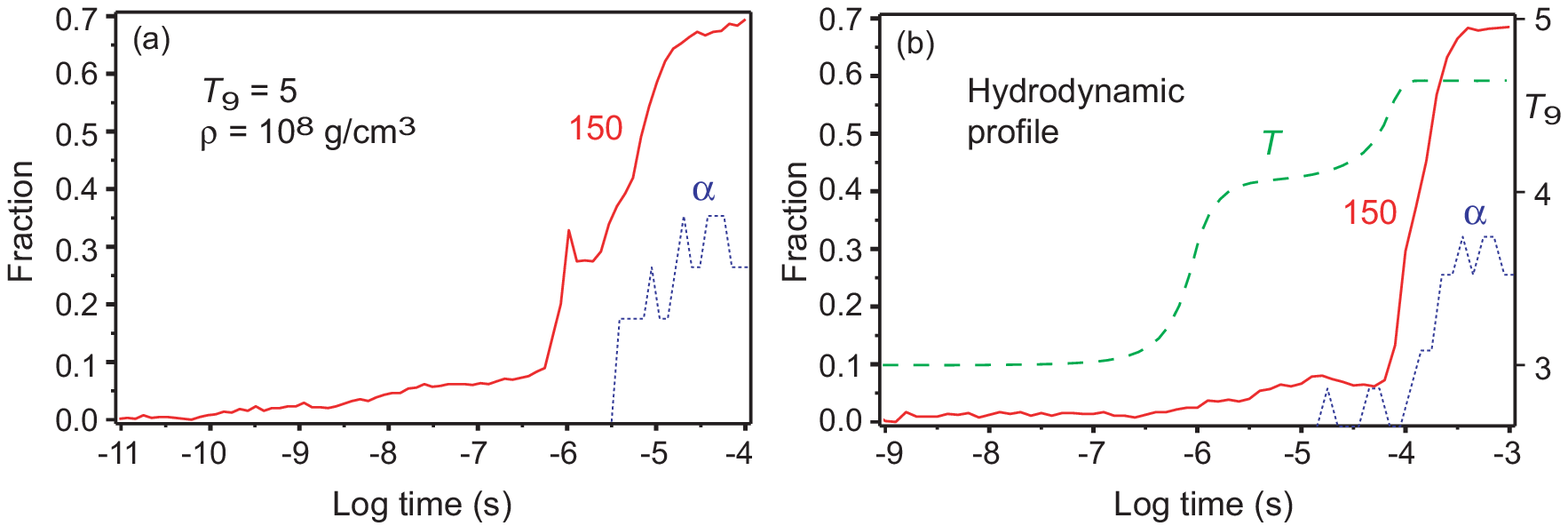}      
{Fraction of equilibrated reactions as a function of time for an alpha network
(dotted blue, labeled $\alpha$) and a 150-isotope network (solid red, labeled
150). (a)~Constant temperature $T_9=5$ and constant
density $\rho = 10^8 \units{g\,cm}^{-3}$ case. (b) Case with hydrodynamic
profile having density essentially constant and the temperature variation given
by the dashed green line, with $T_9$ indicated on the right axis.  In both cases
we see that the 150-isotope network equilibrates much faster and much more
extensively than the alpha network. In these examples we have assumed a reaction
group to be in equilibrium if all isotopes participating in the reactions of the
group have abundances $Y_i$ within 1\% of their equilibrium values.
}
we compare how quickly reactions come into equilibrium (a)~for an alpha network
and for a 150-isotope network under constant temperature and density conditions 
typical of some zones in a Type Ia supernova explosion, and (b)~using a
hydrodynamical profile derived from one zone of a Type Ia supernova simulation.
In this calculation we have not yet implemented the partial equilibrium
approximation, but have integrated using the asymptotic approximation and
tracked the number of reactions that would satisfy the partial equilibrium
condition as a function of time in that case. We see that the effect is quite
pronounced, with the large, realistic network having a much greater fraction of
its reactions satisfying the partial equilibrium condition from very early times
when compared with the alpha network. For example, by the time the alpha network
begins to show any substantial equilibration the 150-isotope network is 30--40\%
equilibrated in both cases. This suggests that the partial equilibrium
approximation may have an even more dramatic effect on the speed with which
large networks can be integrated than we have found for alpha networks.

\subsection{A Toy Model: Protons Plus Alpha-Particles}

Let us illustrate the ideas of the previous section with a toy model that is
simple to understand, but that contains many of the essential
features expected for a realistic network. In \fig{Xcomposite4-isotope_dt}
 \putfig
     {Xcomposite4-isotope_dt}
     {0pt}
     {\figdn}
     {1.0}
     {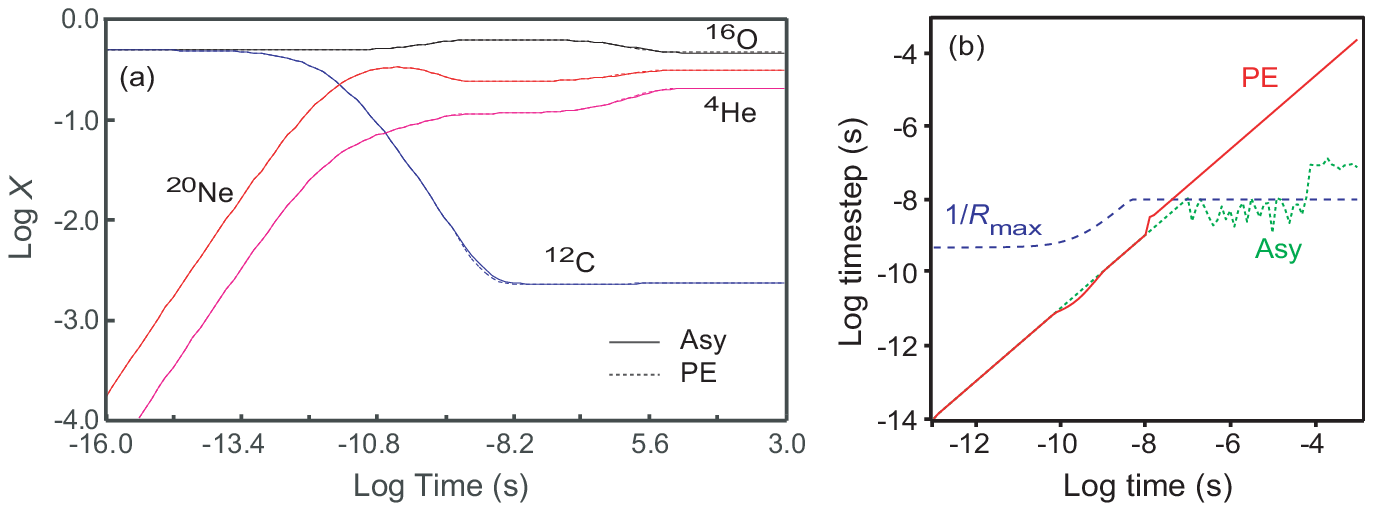}  
{4-isotope alpha network for constant
temperature $T_9 = 5$ and density $\rho = 1\times 10^{8} \units{g\,cm}^{-3}$.
Reactions rates were taken from Ref.~\cite{raus2000} and initially there were
equal mass fractions of \isotope{12}{C} and \isotope{16}{O}. (a)~Mass
fractions (asymptotic solid; asymptotic plus partial equilibrium dotted).
(b)~Timestepping (solid red for asymptotic plus partial equilibrium and dotted
green for asymptotic). The maximum stable purely-explicit timestep is estimated
by $1/R\tsub{max}$ and shown as the dashed blue curve.  }
we display a calculation with a 4-isotope alpha network containing the
isotopes \isotope{4}{He}, \isotope{12}C, \isotope{16}O, and \isotope{20}{Ne},
including all reactions among these species from the REACLIB \cite{raus2000}
library. We see that the partial equilibrium abundances are essentially
identical to those calculated using only the explicit asymptotic algorithm, and
that at the end of the calculation the partial equilibrium calculation is taking
timesteps that are about 3 orders of magnitude larger than those for the
explicit asymptotic method alone, and about 4 orders of magnitude larger than
the maximum timestep that would be stable for a standard explicit method.

In \fig{Xcomposite4-isotopePlus_dt}
 \putfig
     {Xcomposite4-isotopePlus_dt}
     {0pt}
     {\figdn}
     {1.0}
     {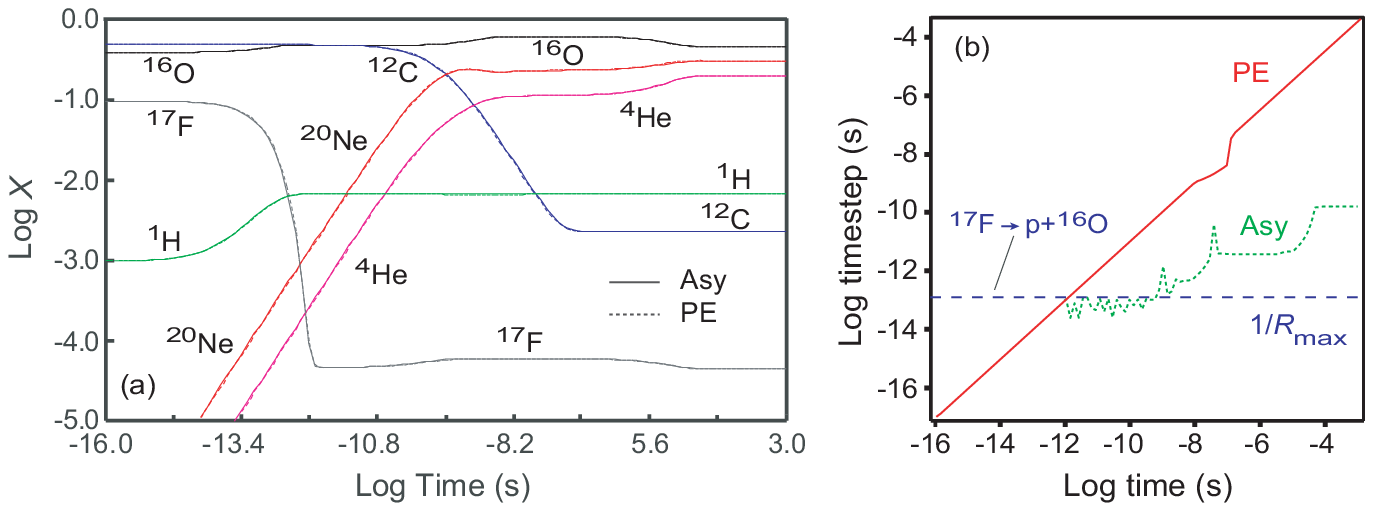}  
{4-isotope alpha network plus protons and \isotope{17}{F}, for constant
temperature $T_9 = 5$ and density $\rho = 1\times 10^{8} \units{g\,cm}^{-3}$.
Reactions rates were taken from Ref.~\cite{raus2000} and the initial mass
fractions 
were $X_0(\isotope{12}{C}, \isotope{16}{O}, \isotope{1}{H}, \isotope{17}{F}) =
(0.500,\, 0.400,\, 0.001, 0.122)$. (a)~Mass
fractions (asymptotic solid; asymptotic plus partial equilibrium dotted).
(b)~Timestepping (solid red for asymptotic plus partial equilibrium and dotted
green for asymptotic). The maximum stable purely-explicit timestep is estimated
by $1/R\tsub{max}$ and shown as the dashed blue line.  It is determined by the
rate for $\isotope{17}{F} \rightarrow p + \isotope{16}{O}$ across the entire
time range shown. This photodisintegration rate is constant because the
temperature is constant.}
we repeat the calculation of \fig{Xcomposite4-isotope_dt}, except that we add to
the 4-isotope alpha network two new isotopes: protons and \isotope{17}{F}. Now
the network contains new reactions involving protons that are much faster than
the alpha reactions and hence is much stiffer. Again we find that the mass
fractions calculated using the asymptotic plus partial equilibrium method and
the asymptotic method alone are almost identical, but the disparity in the
corresponding integration speeds is now much larger. At the end of the
calculation illustrated in \fig{Xcomposite4-isotopePlus_dt}, we see that the PE
integrator is taking timesteps that are about $10^6$ times larger than those for
the explicit asymptotic integrator, and almost $10^9$ times larger than would be
stable with a normal explicit method without the partial equilibrium
approximation. 

This simple example illustrates succinctly the arguments of the previous
section. Comparing \fig{Xcomposite4-isotopePlus_dt} with
\fig{Xcomposite4-isotope_dt}, we see that the addition of a single set of proton
reactions has an enormous influence on the stiffness of the network. In the pure
$\alpha$-particle case of \fig{Xcomposite4-isotope_dt}, the maximum stable
fully-explicit timestep in the approach to equilibrium is of order $10^{-8}
\units s$ and is set by $\alpha$-particle reactions.  In this case a sufficient
number of isotopes become asymptotic to permit an asymptotic timestep of near
$10^{-7}$ seconds in the approach to equilibrium.  This is not a competitive
timestep, since we see from \fig{Xcomposite4-isotope_dt} that the PE method is
able to take timesteps about three orders of magnitude larger, but it is far
better than the situation in \fig{Xcomposite4-isotopePlus_dt}.  There we see
that the maximum stable fully-explicit timestep over the entire range of
integration is of order only $10^{-13} \units s$, because it is set by proton
reactions that are much faster than any $\alpha$-particle reaction, as we have
already illustrated in \fig{maxRateAlphaVs150Hydro}. (Under these conditions the
fastest of these is the photodisintegration $\gamma + \isotope{17}F \rightarrow
p + \isotope{16}{O}$.)

In \fig{Xcomposite4-isotopePlus_dt} the PE method is  taking timesteps in
the approach to equilibrium that are {\em six} orders of magnitude larger than
those of the pure asymptotic method.  The reason for this increased disparity
between the PE plus asymptotic and pure asymptotic methods is not primarily a
better timestep for the PE method (it is taking timesteps in both calculations
that are near the limit imposed by accuracy constraints), but rather the large
depression of the maximum stable asymptotic timestep produced by adding the much
faster proton reactions to the network. However, we see that when the fast
proton reactions are removed from the numerical integration by the PE
approximation, the timestep of \fig{Xcomposite4-isotopePlus_dt} becomes
comparable to that for the much less stiff pure alpha network of
\fig{Xcomposite4-isotope_dt}.

Hence, our toy model supports rather well the speculations of the previous
section. Realistic large networks are much stiffer than $\alpha$-networks, but 
this increased stiffness is associated largely with fast neutron and proton
reactions that may be more susceptible to improvement by partial equilibrium
approximations than simple alpha networks. The  results of this section suggest
that in large realistic networks the systematic removal of the fastest
timescales by the PE approximation may permit timestepping by these explicit
methods that is comparable to that of implicit methods, even when equilibrium is
important.  As we have seen, timestepping comparable to that of implicit methods
in large networks implies that an explicit partial equilibrium code should be
considerably faster than the implicit code, because it can compute each timestep
faster.

\section{\label{sh:summary} Summary and Conclusions}

Numerical integration for extremely stiff reaction networks has typically been
viewed as requiring implicit schemes, because normally explicit timesteps of
efficient size would be unstable.  An alternative to implicit integration
modifies the equations  using approximate solutions to reduce the stiffness, and
then integrates the resulting equations explicitly. For example, asymptotic and
quasi-steady-state methods found a measure of success integrating moderately
stiff chemical-kinetics systems explicitly \cite{oran05,mott00,mott99}, but have
generally failed for extremely stiff systems, giving inaccurate results with
non-competitive integration timesteps. In this paper, and the two preceding it
\cite{guidAsy,guidQSS}, we have presented evidence suggesting that
algebraically-stabilized explicit integration can give correct results and
timesteps competitive  with that of implicit methods for even the stiffest of
networks.

Central to this new view of the efficacy of explicit integration for stiff
systems is the understanding that in reaction networks there are at least three
different sources of stiffness:

\begin{enumerate}

 \item 
{\em Negative populations}, which can occur for an initially small positive
population if an explicit timestep is too large; this anomalous negative
population leads to exponentially growing terms that destabilize the network.

\item
{\em Macroscopic equilibration}, where the right sides of the differential
equations  $dY = F = \Fplus{} - \Fminus{}$ approach a constant derived from the
difference of two large numbers (the total flux in $\Fplus{}$ and total flux out
$\Fminus{}$), and numerical errors in taking this difference destabilize the
network if the timestep is too large. 

\item
{\em Microscopic equilibration}, where on the right sides of
Eq.~(\ref{equilDecomposition}) the net flux in specific forward-reverse reaction
pairs $(f^+_i - f^-_i)$ tends to zero as the system approaches equilibrium; this
leads to large errors if the timestep is too large because the net flux is
derived from the difference of two large numbers and the timescale equilibrating
the populations is short compared with the desired numerical timestep. 

\end{enumerate}
The algebra required to stabilize the
explicit solution depends on which of these forms of stiffness is operative. In
Refs.~\cite{guidAsy,guidQSS} we have shown that asymptotic and
quasi-steady-state approximations give correct results, with timestepping highly
competitive with implicit methods even for the stiffest of thermonuclear
networks, as long as the system exhibits only the first two types of stiffness
listed above. However, in those papers we have also shown that asymptotic and
quasi-steady-state methods give correct results, but with timesteps that are
much larger than for purely explicit methods but still far from competitive with
implicit methods, if there is any significant microscopic equilibration in the
system.

In this paper we have addressed in depth the role of microscopic equilibration
in stiffness and presented a partial equilibrium scheme that can be used to
augment asymptotic and quasi-steady-state methods when equilibrium becomes
important in a network. We have presented strong evidence that this partial
equilibrium scheme can remove the timestepping deficiencies encountered by
asymptotic and quasi-steady-state methods in the approach to equilibrium, making
these methods competitive with implicit methods even near equilibrium. The
methods that we have presented build on earlier work
\cite{oran05,mott00,mott99}, but we find that our versions of asymptotic and
quasi-steady-state methods are far more accurate, and our versions of partial
equilibrium are faster by multiple orders of magnitude, than found in previous
work when applied to extremely stiff astrophysical thermonuclear networks.

The present paper, in conjunction with the previous ones on asymptotic methods
\cite{guidAsy} and quasi-steady-state methods \cite{guidQSS}, suggests that
algebraically-stabilized explicit integration methods are capable of stable
timesteps competitive with those of implicit methods even for extremely-stiff
reaction networks. Explicit methods can compute timesteps faster than implicit
methods in large networks, implying that algebraically-stabilized explicit
algorithms may compete favorably with implicit integration, even in complex,
extremely-stiff applications. Because explicit methods scale linearly with
network size, these findings could permit the coupling of physically more
realistic reaction networks to fluid dynamics simulations in a variety of
fields.

\begin{ack}
We thank Austin Harris for help with some of the calculations, Elisha Feger,
Tony Mezzacappa, Bronson Messer, Suzanne Parete-Koon, and Kenny Roche for useful
discussions, and Suzanne Parete-Koon for a careful reading of the manuscript.
Research was sponsored by the  Office of Nuclear Physics, U.S. Department of
Energy.
\end{ack}

\clearpage
\newpage

\setcounter{section}{0}
\appendix
\section{Appendix:  Reaction Group Classification}
\protect\label{RGclassificationApp}

\noindent
Applying the principles discussed \S\ref{reactionGroupClasses} to the reaction
group classes in Table \ref{tb:reactionGroupClasses} gives the following
systematic partial equilibrium properties of reaction group classes for
astrophysical thermonuclear networks.


\vspace{10pt}

\leftline{{\bf Reaction Group Class A} (a $\rightleftharpoons$ b)}

\parskip = 4pt
\parindent = 0pt

Source term: $\deriv{y_a}{t} = -k\tsub f y_a + k\tsub r y_b\quad$
Constraints:  $y_a + y_b \equiv c_1 = y_a^0 + y_b^0$

Equation: $\deriv{y_a}{t} = by_a + c \quad
b = -k\tsub f \quad c = k\tsub r\quad$
Solution:  $y_a(t) = y_a^0 e^{bt} - \frac cb \left(1-e^{bt}\right)$

Equil.\ solution:  $\bar y_a = -\frac cb = \frac{k\tsub r}{k\tsub f}\quad$
Equil.\ timescale: $\tau = \frac 1b = \frac{1}{k\tsub f}$

Equil.\ tests:  $\frac{|y_i-\bar y_i|}{\bar y_i} < \epsilon_i
\ \  (i = a, b)\quad$
Equil.\ constraint: $\frac{y_a}{y_b} = \frac{k\tsub r}{k\tsub f}$

Other variables:  $y_b = c_1 - y_a $

Progress variable:  $\lambda \equiv y_a^0 - y_a \quad y_a = y_a^0 - \lambda
\quad y_b = y_b^0 + \lambda$

\vspace{15pt}
\leftline{{\bf Reaction Group Class B} (a + b $\rightleftharpoons$ c)}

Source term: $\deriv{y_a}{t} = -k\tsub f y_a y_b + k\tsub r y_c$

Constraints:  
$y_b - y_a \equiv c_1 = y_b^0 - y_a^0\quad$
$y_b + y_c \equiv c_2 = y_b^0 + y_c^0$

Equation: $\deriv{y_a}{t} = ay_a^2 +  by_a + c\quad$
 $a = -k\tsub f \quad b = -(c_1 k\tsub f + k_b)
\quad c = k\tsub r (c_2-c_1)$

Solution: \eq{2body1.4}
$\quad$
Equil.\ solution:  \eq{2body1.6}
$\quad$
Equil.\ timescale: \eq{2body1.8}

Equil.\ tests:  $\frac{|y_i-\bar y_i|}{\bar y_i} < \epsilon_i
\ \  (i = a, b, c)\quad$
Equil.\ constraint: $\frac{y_a y_b}{y_c} = \frac{k\tsub r}{k\tsub f}$

Other variables:  
$y_b = c_1 + y_a  \quad y_c = c_2-y_b $

Progress variable:  $\lambda \equiv y_a^0 - y_a \quad y_a = y_a^0 - \lambda
\quad y_b = y_b^0 - \lambda\quad$
$y_c = y_c^0 + \lambda$

\vspace{15pt}
\leftline{{\bf Reaction Group Class C} (a + b  + c $\rightleftharpoons$ d)}

Source term: $\deriv{y_a}{t} = -k\tsub f y_a y_b y_c+ k\tsub r y_d\quad$
Constraints:
$y_a - y_b \equiv c_1 = y_a^0 - y_b^0$

$\tfrac13 (y_a+y_b+y_c)+y_d \equiv c_3 = \tfrac13 (y_a^0 + y_b^0 + y_c^0) +
y_d^0\quad$
$y_a - y_c \equiv c_2 = y_a^0 - y_c^0$

Equation: $\deriv{y_a}{t} = ay_a^2 +  by_a + c$

$a = -k\tsub fy_a^0 + k\tsub f(c_1+c_2) \quad b = -(k\tsub f c_1 c_2 + k\tsub
r)\quad$
$c = (c_3+\tfrac13 c_1 + \tfrac13 c_2) k\tsub r$

Solution: \eq{2body1.4}
$\quad$
Equil.\ solution:  \eq{2body1.6}
$\quad$
Equil.\ timescale: \eq{2body1.8}

Equil.\ tests:  $\frac{|y_i-\bar y_i|}{\bar y_i} < \epsilon_i
\ \  (i = a, b, c, d)$
$\quad$
Equil.\ constraint: $\frac{y_a y_b y_c}{y_d} = \frac{k\tsub r}{k\tsub f}$

Other variables:  
$y_b = y_a -c_1  \quad y_c = y_a -c_2$
$\quad$
$ y_d = c_3 - y_a + \tfrac13 (c_1 + c_2)$

Progress variable:  $\lambda \equiv y_a^0 - y_a \quad y_a = y_a^0 - \lambda $
$\quad$
$y_b = y_b^0 - \lambda$
$\quad$
$y_c = y_c^0 - \lambda$ 

$y_d = \lambda+y_d^0$

\newpage

\leftline{{\bf Reaction Group Class D} (a + b $\rightleftharpoons$ c + d)}

Source term: $\deriv{y_a}{t} = -k\tsub f y_a y_b + k\tsub r y_c y_d\quad$
Constraints:  
$y_a - y_b \equiv c_1 = y_a^0 - y_b^0$

$y_a + y_c \equiv c_2 = y_a^0 + y_c^0$
$\quad$
$y_a + y_d \equiv c_3 = y_a^0 + y_d^0$

Equation: $\deriv{y_a}{t} = ay_a^2 +  by_a + c$
$\quad$
$a = k\tsub r -k\tsub f \quad b = -k\tsub r (c_2 + c_3)+ k\tsub f c_1
\quad c = k\tsub r c_2 c_3$

Solution: \eq{2body1.4}
$\quad$
Equil.\ solution:  \eq{2body1.6}
$\quad$
Equil.\ timescale: \eq{2body1.8}

Equil.\ tests:  $\frac{|y_i-\bar y_i|}{\bar y_i} < \epsilon_i
\ \  (i = a, b, c, d)$
$\quad$
Equil.\ constraint: $\frac{y_a y_b}{y_c y_d} = \frac{k\tsub r}{k\tsub f}$

Other variables:  
$y_b = y_a - c_1 \quad y_c = c_2-y_a \quad y_d = c_3 -y_a$

Progress variable:  $\lambda \equiv y_a^0 - y_a \quad y_a = y_a^0 - \lambda
\quad y_b = y_b^0 - \lambda$
$y_c = y_c^0 + \lambda$

$y_d = y_d^0 + \lambda$

\vspace{15pt}
\leftline{{\bf Reaction Group Class E} (a + b $\rightleftharpoons$ c + d + e)}

Source term: $\deriv{y_a}{t} = -k\tsub f y_a y_b + k\tsub r y_c y_d y_e$

Constraints:  
$y_a + \tfrac13 (y_c+y_d+y_e) \equiv c_1 = y_a^0 + \tfrac13(y_c^0 + y_d^0 +
y_e^0)$

$y_a-y_b \equiv c_2 = y_a^0 - y_b^0 \quad y_c-y_d \equiv c_3 = y_c^0 - y_d^0$
$\quad$
$y_c - y_e \equiv c_4 = y_c^0 - y_e^0$

Equation: $\deriv{y_a}{t} = ay_a^2 +  by_a + c$

$a = (3c_1 - y_a^0)k\tsub r - k\tsub f \quad b = c_2 k\tsub f - (\alpha\beta
+ \alpha\gamma   + \beta\gamma) k\tsub r$
$\quad$
$ c = k\tsub r \alpha\beta\gamma$

$\alpha \equiv c_1 + \tfrac13(c_3+c_4)
\quad \beta\equiv c_1 - \tfrac23 c_3 + \tfrac13 c_4$
$\quad$
$\gamma \equiv c_1 + \tfrac13 c_3 - \tfrac23 c_4$

Solution: \eq{2body1.4}
$\quad$
Equil.\ solution:  \eq{2body1.6}
$\quad$
Equil.\ timescale: \eq{2body1.8}

Equil.\ tests:  $\frac{|y_i-\bar y_i|}{\bar y_i} < \epsilon_i
\ \  (i = a, b, c, d, e)$
$\quad$
Equil.\ constraint: $\frac{y_a y_b}{y_c y_d y_e} = \frac{k\tsub r}{k\tsub
f}$

Other variables:  
$y_b = y_a - c_2 \quad y_c = \alpha - y_a \quad y_d = \beta - y_a$
$\quad$
$y_e = \gamma-y_a$

Progress variable:  $\lambda \equiv y_a^0 - y_a \quad y_a = y_a^0 - \lambda
\quad y_b = y_b^0 - \lambda\quad$
$y_c = y_c^0 + \lambda$

$y_d = y_d^0 + \lambda \quad y_e = y_e^0 + \lambda$

\vspace{10pt}

In the equilibrium tests we have allowed the possibility of a different
$\epsilon_i$ for each species $i$, but in practice one would often choose the
same small value $\epsilon$ for all $i$. The results presented in this paper
typically have used $\epsilon_i = 0.01$ for all species.

\parindent=4ex
\parskip=0pt

\newpage

\bibliographystyle{unsrt}

\end{document}